\documentclass[aps,pre,nofootinbib,reprint,floatfix,showpacs,amsmath,amssymb]{revtex4-1}
\usepackage{graphicx}% Include figure files
\usepackage{color}
\usepackage{dcolumn}% Align table columns on decimal point
\usepackage{times}
\usepackage[english]{babel}
\usepackage[latin1]{inputenc}
\newcommand{\bfr}{ {\boldsymbol r} }
\newcommand{\ff}{h} %% field variable
\newcommand{\dd}{ {\mathrm d} }
\newcommand{\half}{\frac{1}{2}}
\newcommand{\HH}{\mathcal{H}}
\newcommand{\DD}{\mathcal{D}}

\newcommand{\OO}{\mathcal{O}}
\newcommand{\FF}{F} %% free energy
\newcommand{\hyperF}{\mathcal{F}}
\newcommand{\del}{\partial}
\newcommand{\lbra}{\left\langle}
\newcommand{\rket}{\right\rangle}
\newcommand{\cml}[1]{\lbra #1 \rket_{\rm c}}

\newcommand{\kB}{k_{\rm B}}
\newcommand{\RF}[1]{Eq.~(\ref{#1})}
\newcommand{\zb}{\bar{z}}
\newcommand{\im}{{\rm i}}
\newcommand{\ee}{{\rm e}}
\newcommand{\ie}{{\it i.e.}}
\newcommand{\eg}{{\it e.g.}}
\newcommand{\etc}{{\it etc}} %% notice, no dot at the end
\newcommand{\cf}{{\it c.f.}}

\begin{document}

\title{Effective field theory approach to fluctuation-induced forces between
  colloids at an interface}

\author{Cem~Yolcu, Ira~Z.~Rothstein and Markus~Deserno} 

\affiliation{ Department of Physics, Carnegie Mellon University, 5000 Forbes
  Ave., Pittsburgh, PA 15213, USA }

\date{\today}% It is always \today, today,
% but any date may be explicitly specified


\begin{abstract}
  We discuss an effective field theory (EFT) approach to the
  computation of fluctuation-induced interactions between particles
  bound to a thermally fluctuating fluid surface controlled by surface
  tension. By describing particles as points, EFT avoids computing
  functional integrals subject to difficult constraints. Still, all
  information pertaining to particle size and shape is systematically
  restored by amending the surface Hamiltonian with a derivative
  expansion. The free energy is obtained as a cumulant expansion, for
  which straightforward techniques exist. We derive a {\em complete}
  description for rigid axisymmetric objects, which allows us to
  develop a full asymptotic expansion---in powers of the inverse
  distance---for the pair interaction. We also demonstrate by a few
  examples the efficiency with which multibody interactions can be
  computed. Moreover, although the main advantage of the EFT approach
  lies in explicit computation, we discuss how one can infer certain
  features of cases involving flexible or anisotropic objects. The EFT
  description also permits a systematic computation of ground-state
  surface-mediated interactions, which we illustrate with a few
  examples.
\end{abstract}

\maketitle

\section{Introduction}

Be it the electromagnetic field, the curvature of a surface, or the
composition of an inhomogeneous mixture, boundaries or objects
interacting with a field generally place constraints on the
fluctuations of it. This is known to lead to interactions between the
constraining objects.  The earliest and most famous example of this
effect was discussed in 1948 by Casimir, who showed that two
charge-neutral conducting plates in vacuum attract due to the
constraints they impose on the quantum fluctuations of the
electromagnetic field~\cite{Casimir48, BorMohMosRep, MiltonZeroBook}.
Today it has become customary to name such interactions after Casimir,
even in cases where the underlying fluctuations are not quantum
mechanical but thermal in origin~\cite{KrechBook, KarGolRMP,
  GambassiJPConf}.  In this article, we will investigate a particular
incarnation of this effect, namely forces induced by thermal
fluctuations between particles bound to a fluid surface characterized
by surface tension~\cite{LehOetDieEPL, LehOetPRE, NorOetPRE, LiKarPRL,
  GolGouKarEPL, GolGouKarPRE, YolRotDesEPL}.

The non-trivial aspects of such calculations tend to arise from the
constraints that the extended objects impose on the partition sum.
This issue is usually dealt with by pinning the field to the surface
of the objects through delta-functions in the integration
measure~\cite{LiKarPRL}. A clear exposition of this method, applied to
compact objects in fluid membranes and films, can be found in
Ref.~\cite{GolGouKarEPL}.  This method was later refined in the
context of the electromagnetic Casimir
effect~\cite{EmiGraJafKarPRD}. In the latter work, the constraints of
the objects enter the interaction energy through their scattering
matrix coefficients. In fact, usage of scattering approaches for
electromagnetic Casimir interactions has a long history and has lead
to some recent developments~\cite{BordagPRD, BulMagWirPRD, MilWagPRD,
  EmiGraJafKarPRL} in the field, concerning analytical results. Very
recently, thermal Casimir interactions between biological membrane
inclusions were studied in a similar spirit~\cite{LinZanMohPryPRL}.

In this paper we employ a different strategy to streamline the
boundary condition issue, namely, reverting to a point particle
description. We emphasize right from the start that this is not an
uncontrolled approximation for the shape of the extended objects,
because we can systematically rescue the complete finite size
information by means of effective field theory (EFT). The key
philosophy behind EFT is separation of scales, which is deeply rooted
in the concept of renormalization (for a review, see
Ref.~\cite{TASI}). Originally developed in the context of quantum
field theory, it has been applied only quite recently in a purely
classical framework by Goldberger \emph{et al}.~\cite{GolRotPRD}, who
used it to study the gravity wave profile for inspiralling black
holes. This particular incarnation of EFT has subsequently been
utilized to derive not only new results in gravitational wave
physics~\cite{[{See, for example }][{, and references
      therein}]PorRosRot} but also to calculate the leading order
finite size correction to the Abraham-Dirac-Lorentz radiation reaction
force law in classical electrodynamics~\cite{GalLeiRotPRL}. Both of
these applications dealt with intricate boundary condition problems
for classical non-fluctuating fields, whereas the present article
extends the formalism to allow for finite temperatures. We recently
introduced this generalization in Ref.~\cite{YolRotDesEPL}; here we
develop the formalism in greater detail, expand on our previous
results, and discuss further physical situations.

Examples of point particle approaches employed to compute
surface-mediated forces have appeared several times in the
past~\cite{DomFouEPL, Netz97, MarMisEPJE}, in the context of
biological membranes. However, in most of these references the final
answer depends on the short wavelength cutoff of the continuum theory,
which is ambiguous by construction. To alleviate the issue, the cutoff
is usually fixed by reinterpreting it as a multiple of the inclusion
size, but there are at least two problems with this: $(i)$ The size of
the objects embedded in the field and the short wavelength cutoff of
the theory are generally unrelated, and $(ii)$ even if the objects do
have sizes of the same order as the cutoff, it is not clear how to
treat particles with different radii. Instead, the proper way to free
physical quantities from dependence on some ambiguous cutoff is via
renormalization techniques. Our approach is constructed such that the
cutoff never ``contaminates'' the interactions and therefore never
inappropriately restricts particle sizes. Another difference between
earlier point-particle approaches and ours is the fact that our
treatment is not approximate in that all finite size information is
systematically retained in the point particle picture. Unless this is
done, it would not even make sense to attempt calculating higher order
corrections to asymptotically correct results.

For the sake of clarity, we restrict in this paper to the case of
rigid particles, \eg\ colloids adsorbed at a fluid-fluid interface. In
other words, within the region of the surface occupied by the
particle, fluctuations are not only reduced, but are {\em frozen}, in
a manner compatible with the boundary conditions at the circumference
of the particle and its permissible rigid body motions. Furthermore,
we will assume the shape of the particles are such that they have a
circular footprint on the surface. While these two assumptions allow
us to carry out a more transparent discussion of the EFT formalism, we
nonetheless outline what is entailed in relaxing these assumptions.
We also briefly turn our attention to interactions that are not
induced by fluctuations but by permanent shape-irregularities of the
particles. The formalism allows a systematic approach to calculate
these interactions without resorting to a superposition assumption to
disentangle the boundary conditions of the particles from one another.

In the following section of our article, we begin our discussion by
reviewing the nature of the surface and fluctuation-induced
interactions we are interested in, as well as outlining our effective
field theory approach.  We then proceed in Section~\ref{sec:EFT} with
constructing the effective theory and computing interactions in an
asymptotic expansion. First, we illustrate the computation by
considering only one of the many terms in the expansion as an example
in Section~\ref{sec:Casimir}. Afterwards, we derive the interaction
free energy of a pair of particles as a full expansion, \ie\ valid at
arbitrary separation, and extract the leading near-contact
behaviour. We then compute the leading non-trivial triplet and
quadruplet interactions, as well as briefly discuss the case of soft
or non-axisymmetric particles. Finally, we show how the formalism can
be used for problems where the particles impart a permanent
deformation on the surface.  To separate the main arguments from
mundane technicalities, we have collected details of certain
calculations as well as some auxiliary discussions in four appendices.

\section{Formalism}

\subsection{The surface}

We are interested in a fluid surface whose behaviour is determined by
the surface tension energy $ \sigma \int \dd A$, where the integral
encompasses the entire surface and $\sigma$ is the surface tension. A
nearly flat surface can be parameterized using the Monge
gauge~\cite{SafranBook}, describing the shape in terms of its
orthogonal deviation $\ff$ from a flat $xy$ base plane. This yields
the energy functional
\begin{equation}
  \HH_{\rm surf} [\ff] = \sigma \int\limits_{\mathcal{S}} \dd^2r
  \sqrt{ 1+ \ff_{i}^2 (\bfr) } \ ,
\end{equation}
where we write the gradient with the partial derivative subscript $i$.
Note that the domain $\mathcal{S}$ of the integral is not the whole
$\mathbb{R}^2$ but excludes the regions defined by the projection of
the colloids onto the base plane. The presence of the particles
imposes conditions on $\ff$ at the boundaries of these regions. For
concreteness, let us picture circular disks floating on the surface,
which dictate the height of the fluid film anywhere along their
circumference.

Under the further assumption of a weakly-deformed surface, $ \lvert
\nabla \ff \rvert \ll 1$, one finally arrives at the quadratic surface
Hamiltonian
\begin{equation}
  \HH_{\rm surf} [\ff] = \half \sigma \int \limits_{\mathcal{S}}
  \dd^2 r \, \ff_i^2 (\bfr) \ , \label{eq:H}
\end{equation}
up to an irrelevant constant. The existence of this integral defines
the function space $\ff (\bfr)$ belongs to.

\subsection{Fluctuation-induced interactions}
 
At a given temperature, the surface will fluctuate around its ground
state shape, which minimizes the energy as defined in \RF{eq:H}, while
still subject to the boundary conditions imposed by the particles,
\ie\ the continuity of the surface along their circumference.  These
         {\em local} constraints on the fluctuations will result in
         the partition function and the free energy being dependent on
         the spatial arrangement of the constrained regions, hence
         giving rise to forces.

Formally, the free energy is given as
\begin{equation}
  \beta \FF = -\log \int \DD \ff \, \ee^{ -\beta \HH_{\rm surf} [\ff]}
  \ , \label{eq:pf}
\end{equation}
where the partition function is a functional integral over all {\em
  permissible} field configurations $\ff (\bfr)$, \ie\ those
compatible with the boundary conditions imposed by the particles.
This integral is a Gaussian integral, on account of the Hamiltonian
(\ref{eq:H}) being quadratic in the field, but its evaluation is not
straightforward. Though one can formally write the free energy
(\ref{eq:pf}) as $ \sim \log \det \mathbb{K}$, this does not help with
explicit computation, as the kernel $\mathbb{K} = -\nabla^2$ (see
Eqs.~ (\ref{eq:H}) and (\ref{eq:pf})) is hard to diagonalize in this
{\em constrained} function space~\footnote{The integration kernel in
  \RF{eq:H} involves a Laplacian operator that acts on functions
  defined only over a subset $\mathcal{S}$ of $\mathbb{R}^2$.
  Integrating by parts, \RF{eq:H} can be massaged into $ \frac
  {\sigma} {2} \int_{\mathcal{S}} \dd^2 r \, \ff (-\nabla^2) \ff +
  \frac {\sigma} {2} \int_{\del \mathcal{S}} \dd \ell \, \hat{n}_i \ff
  \del_i \ff $ which does not have the form $ \half \int \dd^2 r \,
  \dd^2 r' \, \ff(\bfr) K(\bfr, \bfr') \ff(\bfr') $ with a
  translationally invariant kernel $K(\bfr, \bfr') = K(\bfr - \bfr')$
  which could be diagonalized easily in momentum space.}. One wishes
to free the functional integral in \RF{eq:pf} from the finite sized
regions of constraint, such that the integral becomes a
``straightforward'' Gaussian.

One possible means to this end involves extending the validity of the
Hamiltonian (\ref{eq:H}) to the entirety of $\mathbb{R}^2$ at the
expense of modifying the integration measure by delta functions,
appropriately chosen to make sure the functional integral in
\RF{eq:pf} only sifts out permissible field shapes
$\ff(\bfr)$~\cite{LiKarPRL, LehOetDieEPL, LehOetPRE, NorOetPRE,
  GolGouKarEPL, GouBruPinEPL, *[{erratum: ibid.\ }][]GouBruPinEPLe,
  DomFouEPL, EmiGraJafKarPRD}. In this article, we follow another
path, which encodes the constraints in {\em additional terms} to the
free surface Hamiltonian. This is different from the approach taken in
Refs.~\cite{GouBruPinEPL, ParLub} in that it is {\em not} necessary to
assume these additional terms to be {\em small}, which corresponds to
the case when the constraints on the fluctuation of the particles are
not rigid. Even when one deals with rigid constraints, one can compute
the free energy in a convergent perturbation expansion. This is
achieved by means of an effective field theory (EFT).

\subsection{The effective theory} 

As we have alluded to before, we want to describe the system with an
effective Hamiltonian such that we can write the partition function as
$Z = \int \DD \ff \, \ee^{-\beta \HH_{\rm eff} [\ff]}$ with the field
variations {\em unconstrained}. This allows the evaluation of the
integral over a Gaussian measure with the familiar translationally
invariant harmonic kernel.  To this end, we construct the effective
theory as $\HH_{\rm eff} = \HH + \Delta \HH$ where $\HH$ is the free
surface Hamiltonian
\begin{equation} \label{eq:Hfree}
  \HH [\ff] = \half \sigma \int \limits_{\mathbb{R}^2} \dd^2r
  \, \ff^2_i (\bfr)
\end{equation}
and the perturbation $\Delta\HH$ is a collection of {\em local} terms
specifically designed to capture the constraints, as we will discuss
shortly. In this form, the effective theory describes a system of
point-like particles embedded in an otherwise homogeneous
surface. 

The free energy, measured with respect to the free energy of the
unperturbed surface, is then calculated in a cumulant expansion
\begin{align} 
  \beta \FF &= -\log \int \DD \ff \, \ee^{-\beta \HH} \ee^{-\beta
    \Delta\HH} \label{eq:pfeff} \\ &= -\log \lbra \ee^{-\beta \Delta
    \HH} \rket = -\sum_{q=1} ^{\infty} \frac{1} {q !}  \cml{ \left(
    -\beta \Delta \HH \right)^q} \ .\label{eq:cumex}
\end{align}
The last step above can be taken as the definition of the cumulants
denoted as $\cml{\ldots}$, which involve many-point connected
correlation funtions of the field. Since averages are computed in the
Gaussian ensemble determined by the free Hamiltonian $\HH$, the
many-point correlators break up into a product of two-point
correlators
\begin{equation}\label{eq:twopt}
  \lbra \ff(\bfr) \ff(\bfr') \rket = \frac{1} {\beta \sigma} G (\bfr,
  \bfr') = -\frac {1} {4\pi \beta \sigma} \log (\bfr -\bfr')^2 \ ,
\end{equation}
due to Wick's theorem. Here, $G(\bfr, \bfr')$ is the Green function or
inverse of operator $-\nabla^2$, \ie\ $-\nabla^2 G(\bfr, \bfr') =
\delta (\bfr -\bfr')$.

The aim is that the effective theory reproduces the correct free
energy; that of the original system with finite sized regions of
constraint, or the {\em full} theory as we will henceforth refer to
it.  This is ensured by matching a set of physical observables in the
two theories, a procedure which will be explored in detail further in
Section~\ref{sec:matching}.

In general, $\Delta \HH$ consists of all polynomials in the
derivatives of the field $\ff$ consistent with the symmetries of the
problem, evaluated at the position of each particle. As such, it has
the form of an operator product expansion~\cite{Cardy} involving terms
like $ A_a \ff_{i}^2 (\bfr_a)$, $B_a \ff_{ij}^2 (\bfr_a)$, $C_a
\ff_{ij}^2 (\bfr_a) \ff_i^2 (\bfr_a)$ {\em etc}., where the label $a$
denotes a particle and $\bfr_a$ its appropriately defined position,
\eg\ the center of a circular disk.  The prefactors $A$, $B$,
$C$, \ldots are called Wilson coefficients and have to be fixed by the
aforementioned matching procedure. An expansion of similar spirit was
employed in Ref.~\cite{BurEisPRL} for studying interactions
involving spherical objects in a critical fluid.

Let us briefly discuss the temperature dependence of the free energy
expansion (\ref{eq:cumex}). Every cumulant, being the thermal average
of a power of $\beta \Delta \HH$, will involve many-point correlators
that consist of a multiplicative combination of terms in $\Delta
\HH$. Recalling from \RF{eq:twopt} that every pair of field
occurrences in these many-point correlators carries a factor
$\beta^{-1} = \kB T$, each quadratic term in $\beta \Delta \HH$ can be
counted as a factor $\beta^0$, each quartic term as $\beta^{-1}$, and
so on, in the term of expansion (\ref{eq:cumex}) they appear in. From
this we see that, the traditionally encountered form $\FF = \kB T
f\left( \{ \bfr_a \} \right)$ of fluctuation-induced
interactions~\cite{LiKarPRL, LehOetDieEPL, LehOetPRE, NorOetPRE,
  GolGouKarEPL, GouBruPinEPL, ParLub, DomFouEPL} stems from quadratic
terms in $\Delta \HH$. This allows the interpretation of the
interactions as those between {\em induced} ``capillary charges,''
arising from thermal fluctuations around each particle and subsequent
polarizations due to the deformation that propagates from
those. Similarly, one can see that higher-than-quadratic terms in
$\Delta \HH$ give rise to an excess of factors of $\kB T$ in the
cumulants in \RF{eq:cumex}. Therefore, such terms in the effective
theory produce a free energy $\FF \sim (\kB T)^{p \ge 2} $. These are
higher order fluctuation corrections to the interaction.  However, to
treat these {\em non-linear} corrections properly, one has to also
consistently relax the weak deformation assumption on the free surface
Hamiltonian (\ref{eq:H}). This involves quartic (or higher order)
terms in the free Hamiltonian that can also be treated
perturbatively. While such extensions do not affect the basic EFT
formalism, they require additional field theoretical sophistication
that would distract from the basic idea we wish to communicate here,
and hence we defer such non-linear corrections to future work.

\section{EFT of flat disks on a film} \label{sec:EFT} 

After briefly outlining the main idea of our formalism, we will now
proceed by applying it to a specific problem. We want to compute the
fluctuation-induced interaction between colloids with a circular
footprint on the base plane. For this, we write the effective theory
as $\Delta \HH= \sum_a \Delta \HH_a$, where $a$ labels the
particles. For one particle located at $\bfr_a$, $\Delta \HH_a$ is a
{\em derivative expansion} of the form
\begin{equation} \Delta \HH_a = \half \left.\! \left( C_a^{(0)} \ff^2
      +C_a^{(1)} \ff_i^2 + C_a^{(2)} \ff_{ij}^2 + \ldots \right)
      \right\vert_{\bfr=\bfr_a} \label{eq:drvex} \ ,
\end{equation}
encoding a {\em quadratic} excess energy for deforming the field in
the vicinity of that particle. Denoting the radius of disk $a$ by
$R_a$, dimensional analysis shows that each Wilson coefficient
$C_a^{(\ell)}$ scales as $\sigma R_a^{2\ell}$, such that the limit
$R_a \rightarrow 0$ is well-defined. It is also worth pointing out
that, applying a Hubbard-Stratonovich transformation to the partition
function in \RF{eq:pfeff}, the effective Hamiltonian $\HH_{\rm eff} =
\HH + \sum_a \Delta \HH_a$ can be brought into the form of the
effective action of Emig {\em et al.}~\cite{EmiGraJafKarPRD}

As we mentioned earlier, the terms in \RF{eq:drvex} represent induced
charges localized at the points $\bfr_a$. Thus, the coefficients
$C_a^{(\ell)}$ will be referred to as {\em polarizabilities}. One can
see this clearly by considering a generic term in \RF{eq:drvex} around
an applied or background field. Choosing $\bfr_a=0$ and dropping the
label $a$ for convenience, one substitutes $ \ff \rightarrow \ff^{\rm
  bg} + \ff $ into the term $ (1/2) C^{(\ell)} \left( \del^\ell \ff
\right)^2$ and observes the term linear in $\ff$:
\begin{align}
  C^{(\ell)} \del^\ell \ff^{\rm bg}(0) \del^\ell \ff (0) =& \int \dd^2
  r \, \ff (\bfr) \nonumber \\ & \times C^{(\ell)} \del^\ell \ff^{\rm
    bg} (0) \, (-\del)^\ell \delta(\bfr).
\end{align}
One can then identify the localized source
\begin{equation} \label{eq:source}
  J (\bfr) = (-)^\ell C^{(\ell)} \del^\ell \ff^{\rm bg} (0) \,
  \del^\ell \delta(\bfr) \ .
\end{equation}
This source is an $\ell^{\rm th}$ order multipole induced by the
$\ell^{\rm th}$ derivative of the background, created by thermal
fluctuations or the field of another such charge.

The constraints imposed on the motion of the particle (typically by
the experimental situation at hand) determine whether a certain
multipole charge can be induced or not. For instance, if the particle
is free to move vertically, there will be no monopole induced, or
$C^{(0)} =0$, because the particle can accommodate a nonzero $\ff (0)$
by moving up or down, without an excess energy cost. Similarly, the
dipole polarizability $C^{(1)}$ vanishes if there are no constraints
on the tilt degree of freedom of the particle, as it can then adjust
freely to a nonzero $ \del_i \ff (0)$. We will reprise these arguments
in due time when we discuss the leading order interaction between the
particles.

\begin{figure}[!tb]
  \includegraphics[scale=.7] {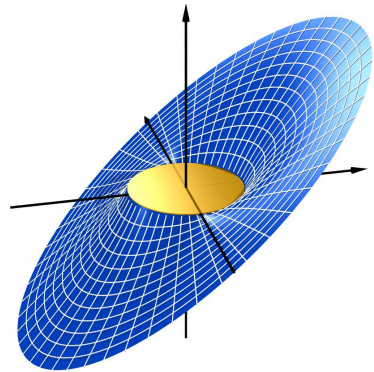} \caption{ The figure
    illustrates how a rigid, horizontally fixed particle of radius $R$
    creates a local field deformation when exposed to a tilted planar
    background, $\ff^{\rm bg} = r \cos \varphi$. The total deformation
    is the sum of the background and the induced deformation, $ \delta
    \ff = - (R^2/r) \cos \varphi$. The discontinuity of the (normal)
    gradient of the total shape along the circumference of the
    particle is a dipole-like ($\sim \cos \varphi$) ``capillary
    polarization'' that can be thought of as sourcing the response
    $\delta \ff$.} \label{fig:polarization}
\end{figure}

The presence of a multipole like \RF{eq:source} on the surface creates
a deformation around it, which is simply the following convolution of
the charge with the Green function:
\begin{align}
   \delta \ff (\bfr) =& - \frac{1} {\sigma} \int \dd^2 r' J(\bfr') G(\bfr',
   \bfr) \nonumber \\ =& - \frac{C^{(\ell)}} {\sigma} \del^\ell
   \ff^{\rm bg} (0) \, \del^\ell G(0, \bfr) \ . \label{eq:response}
\end{align}
For example, Fig.~\ref{fig:polarization} depicts the response of a
horizontally fixed disk to a dipole background, $\ff^{\rm bg} = r \cos
\varphi$. The fluctuation-induced interaction can be viewed as the
interaction between these induced multipoles mediated by the surface.

Notice that the multipole responses of order $\ell > 0$ decay as
$r^{-\ell}$, whereas the monopole response does not. In fact, the
monopole response is not even square integrable, \ie\ its energy
\RF{eq:H} diverges logarithmically. However, this is not a property of
the EFT point particle description, but of the surface Hamiltonian
itself. Therefore, calculating interactions between objects with
nontrivial monopole responses in $\mathbb{R}^2$, \ie\ disks with
pinned height, is intrinsically an ill-defined problem and requires
regularization of the surface Hamiltonian---whether one is using EFT
or not. The issue can be alleviated by considering a finite subset of
$\mathbb{R}^2$ or modifying the theory with a damping term. However,
since this would distract from the main idea we want to communicate in
this article, we will defer this case to Appendix~\ref{sec:monopole}
and avoid monopoles in the main text by restricting to particles that
can fluctuate up and down freely, \ie\ particles for which $C^{(0)} =
0$.

Having discussed the multipole responses, one more note for
completeness is in order. Although one must build $\Delta \HH$ from
all scalars that the symmetries of the problem allow, terms involving
repeated indices, such as $ \ff_{ii}^2$, $\ff_{iij}^2$, \etc.\ are
redundant. One way to see this is that the response \RF{eq:response}
of such a charge to a background would necessarily be {\em local}, as
$ G_{ii} (0, \bfr) \equiv \delta (\bfr)$, and hence such charges do
not interact with anything away from them.

\subsection{Fixing the polarizabilities: matching} \label{sec:matching}

Eq.~(\ref{eq:drvex}) is the most general quadratic Hamiltonian
compatible with the circular symmetry of one particle.  The only thing
we have not established yet are the exact values of the polarizability
coefficients, which we will attend to in this section. These values
are fixed by matching a set of observables in the effective (point
particle) and full (finite sized particle) theories. Although any
suitably chosen set of observables will do, we will consider the most
obvious one here: the polarization response to an applied field. If
one is dealing with more complicated physics and boundary conditions
or shapes, a different choice of observable may be more practical
(\eg\ for numerical matching) or even necessary.

We gave the general expression for an $\ell^{\rm th}$ order induced
multipole response in \RF{eq:response}. Clearly, the background that
will induce this response is one that is $\ell^{\rm th}$ order in the
coordinates, and the deformation decays like $ \del^\ell G(0, \bfr)
\sim r^{-\ell}$. The polarizability $C^{(\ell)}$ should be fixed such
that this EFT response is identical to the response in the full theory
of the particle to the same background. Given the correspondence
between our present Hamiltonian and classical electrostatics in 2D,
the latter response is found analogously to the problem of a conductor
in an external electric field.

If we write the derivatives in \RF{eq:response} explicitly, this
expression involves the index contraction $\ff^{\rm bg}_{i_1 \ldots
  i_\ell} (0) G_{i_1 \ldots i_\ell} (0, \bfr)$ between partial
derivatives of the incident background and the Green function.  The
simplest choice of background is $ \ff^{\rm bg} = \alpha r^\ell \cos
\ell \varphi$. While a cartesian index contraction is not hard to do
when the polarization is of order 1 or 2, a general expression is
difficult to come by in this way for high orders. Luckily, if one
carries out the contraction in complex coordinates $(z, \bar{z})$
instead of $(x,y)$, a general expression is found easily. The result
is that the effective object will respond to a background of the form
$ \ff^{\rm bg} = \alpha r^\ell \cos \ell \varphi$ by creating the
deformation
\begin{equation} \label{eq:effresp} 
  \delta \ff (\bfr) = - C^{(\ell)} \frac {2^\ell \ell! (\ell-1)!
    \alpha} {4 \pi \sigma} \frac {\cos \ell \varphi} {r^\ell} \ ,
\end{equation}
which is derived in Appendix~\ref{sec:complex}.

In the full theory, one has to solve the boundary value problem
$\nabla^2 \delta \ff =0$ for $r > R$ with the boundary condition $
\left. \ff \right | _{r=R} = \left. (\ff^{\rm bg} + \delta \ff) \right
\vert_{r=R} = 0$~\footnote{Strictly speaking, the boundary condition
  is $ \left. \ff \right | _{r=R} = a + {\boldsymbol b} \cdot \bfr$
  where $a$ and ${\boldsymbol b}$ are free parameters describing the
  height and tilt (if applicable) fluctuations of the colloid. Their
  function is to satisfy the boundary condition readily without the
  need for an induced response for those backgrounds that the colloid
  can align with.}. The full set of harmonic functions with
square-integrable gradient has the form $r^{-\ell} \cos \ell \varphi$
with $\ell \in \{1,2,\ldots\}$. Hence the solution in the case of a
background $\ff^{\rm bg} = \alpha r^\ell \cos \ell \varphi$ is found
to be
\begin{equation} \label{eq:fullresp}
  \delta \ff (\bfr) = -\alpha R^{2\ell} \frac{ \cos \ell \varphi}
         {r^\ell} \ .
\end{equation}
Comparison of Eqs.~(\ref{eq:effresp}) and (\ref{eq:fullresp}) yields
the complete set of polarizabilities
\begin{equation} \label{eq:C} C^{(\ell)} = \frac{ 4 \pi R^{2\ell} \sigma}
  {2^\ell \ell! (\ell-1)!} \quad , \quad \ell=1,2,3\ldots \ ,
\end{equation}
for one object. For the sake of clarity, let us rewrite the complete
$\Delta \HH$ by restoring the particle labels $a$:
\begin{equation} \label{eq:DeltaH} 
  \Delta \HH [\ff] = \half \sum_a \sum_{\ell=1} ^{\infty} C^{(\ell)}_a
  \left[ \del^\ell \ff (\bfr_a) \right]^2 \ ,
\end{equation}
where $\del^\ell$ denotes $\del_{i_1} \del_{i_2} \ldots
\del_{i_\ell}$. We now know the point-particle EFT Hamiltonian which
--- to quadratic order in the field---is rigorously equivalent to
the full theory.

As stated earlier, we do not intend to address higher-than-quadratic
terms in the effective theory here. However, we would like to stress
that the matching procedure is essentially the same even if such terms
are considered.  The steps taken since the beginning of
Sec.~\ref{sec:EFT} do not {\em rely} on the responses being linear.
What has been presented boils down to computing the one-point function
$\lbra \delta \ff(\bfr) \rket$ of the response $\delta \ff(\bfr)$ of
one isolated object to a background in the effective (point particle)
and full theories.  Generally, one matches such $n$-point functions,
or other observables that are functions or functionals of them, across
both theories to fix the Wilson coefficients.  For the linear case, it
turned out to be very straightforward to fix the coefficients
independently of each other by matching the aforementioned one-point
function.  If nonlinearities are also accounted for, one may need to
match more $n$-point functions to obtain a sufficient number of
linearly independent matching conditions to fix the values of the
coefficients and there may not be a simple choice of background that
automatically decouples them.  However, despite the tedium introduced
by the nonlinearities, the procedure is not intrinsically problematic.

\subsection{Fluctuation-induced interactions}\label{sec:Casimir}

\begin{figure}[!bt]
  \includegraphics[scale=.75]{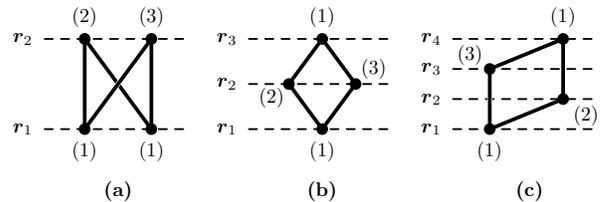} \caption{Diagrams depicting pair
    (a), triplet (b) and quadruplet (c) interactions stemming from the
    $4^{\rm th}$ cumulant. Dashed lines represent the {\em
      world-lines} of particles and have been labeled by the
    particle. The number in parentheses at each vertex shows the
    multipole order at the associated point in the scattering
    process. Each propagator is subject to $\ell_1 + \ell_2$ partial
    derivatives where $\ell_1$ and $\ell_2$ are the multipole orders
    at the vertices it connects. Consequently, all these diagrams are
    interactions of order $2(1+2+3+1)=14$ in the inverse
    inter-particle separations. Also note that we show only one of
    three possible 4 body connections in (c), as our aim at this stage
    is just to work an example and not to be complete.
  } \label{fig:4}
\end{figure}

With the effective theory established, we can now turn our attention
to evaluating the cumulant expansion (\ref{eq:cumex}) explicitly.
Taking powers of \RF{eq:DeltaH} one observes that the $q^{\rm th}$
cumulant in \RF{eq:cumex} involves $2q$-point connected correlation
functions. This can be represented as connected Feynman diagrams with
$q$ vertices and 2 links coming out of each vertex (see
Fig.~\ref{fig:4}, for instance), as one can see by Wick contraction of
the $2q$ point correlator. Every vertex contributes a factor of $\beta
C^{(\ell)}/2$ while every link affords a two-point correlator (or
propagator) as in \RF{eq:twopt}, acted upon by derivatives stemming
from \RF{eq:DeltaH}.  Recalling that $C^{(\ell)} \sim \sigma$, one can
see that there will be no factors of $\beta$ or $\sigma$ left in any
cumulant, \ie\ the dimensionless free energy $\beta \FF$ is purely a
function of the spatial configuration of particles. Each correlator
$\lbra \del^\ell \ff (\bfr_a) \del^m \ff (\bfr_b) \rket \sim \del^\ell
\del^m G(\bfr_a, \bfr_b)$ brings a factor $\vert \bfr_a -\bfr_b
\vert^{-(\ell+m)}$. Consequently, an interaction among $q$ multipoles
of orders $\ell_1$ to $\ell_q$ will scale with a total of $-2(\ell_1 +
\ell_2 + \ldots + \ell_q)$ powers of the separations between
them. Accordingly, one knows how many multipole orders and how many
cumulants must be retained, in order to achieve any given level of
accuracy of an asymptotic expansion of the free energy, expressed in
inverse powers of inter-particle distances.

\pagebreak
It is instructive to consider one example term from the cumulant
expansion, to make these ideas clearer. The fourth cumulant will serve
well since it contains pair, three body, and four body interactions,
while the expressions are still not very cluttered. Let us first write
down the fourth cumulant directly from Eqs.~(\ref{eq:cumex}) and
(\ref{eq:DeltaH}):
\begin{widetext}
\begin{equation} 
  \beta \FF^{(4)} \mathrel{\mathop:}= - \frac{\beta^4} {4!}
  \sum_{\substack{a,b\\c,d}}
  \sum_{\substack{\ell_a,\ell_b\\ \ell_c,\ell_d}}
  \frac{C^{(\ell_a)}_{a \!\!\! \phantom{l}} C^{(\ell_b)}_b
    C^{(\ell_c)}_{c \!\!\! \phantom{l}} C^{(\ell_d)}_d} {2^4} \lbra
  \left[ \del^{\ell_a} \ff (\bfr_a) \right]^2 \left[ \del^{\ell_b} \ff
    (\bfr_b) \right]^2 \left[ \del^{\ell_c} \ff (\bfr_c) \right]^2
  \left[ \del^{\ell_d} \ff (\bfr_d) \right]^2 \rket_{\rm
    c}. \label{eq:4}
\end{equation}
\end{widetext}
The summations are over the particles and the multipoles on each. In
this section, we will discuss only a few terms embodied in these sums,
chosen as instructive examples (shown in Fig.~\ref{fig:4}).

We will begin by considering the pair interaction between particles 1
and 2, depicted in Fig.~\ref{fig:4}(a). There are two factors one
needs to compute: $(i)$ the number of times the relevant product of
polarizabilities, namely $ C^{(1)}_1 C^{(1)}_1 C^{(2)}_2 C^{(3)}_2 $,
occurs in the sum (\ref{eq:4}), and $(ii)$ the multiplicity of the
Wick contraction of the 8-point connected correlation function. The
first of these factors is simply the multinomial coefficient $
\binom{4}{2,1,1} = \frac{4!}{2! 1! 1!}$, since what we are looking for
is the factor in front of $a^2bc$ in the expansion of
$(a+b+c+\ldots)^4$. Next, the multiplicity of the Wick contraction, or
the diagram Fig.~\ref{fig:4}(a), is $2^4$ due to the possibility of
flipping the legs of each vertex.\footnote{It is generally $2^q$ with $q$
being the cumulant order, except for $q=2$, where it is $2$.} Thus the
pair interaction of Fig.~\ref{fig:4}(a) can be written as
\begin{align} 
  \beta \FF (\text {\small Fig.~\ref{fig:4}(a)}) = & -\frac{1} {4!}
  \binom{4}{2,1,1} \frac { {C_1^{(1)}}^2 C_2^{(2)} C_2^{(3)}} {(2
    \sigma)^4} \nonumber \\ & \times 2^4 G^{12}_{ijk} G^{21}_{jkl}
  G^{12}_{lmnp} G^{21}_{mnpi} \ , \label{eq:pair4}
\end{align} 
where the shorthand $G^{ab} = G(\bfr_a,\bfr_b)$ has been
introduced. For completeness, we also present the result after doing
the index contraction (see Appendix~\ref{sec:complex}) and
substituting polarizabilities from \RF{eq:C}:
\begin{equation}
  \beta \FF (\text{\small Fig.~\ref{fig:4}(a)}) = -6 \frac{ R_1^4
    R_2^{10}} {r_{12}^{14}} \ ,
\end{equation}
where $r_{ab} = \vert \bfr_a -\bfr_b \vert$.

The triplet interaction depicted in Fig.~\ref{fig:4}(b) requires
consideration of terms in \RF{eq:4} which involve the polarizabilities
$C^{(1)}_1$, $C^{(2)}_2$, $C^{(3)}_2$ and $C^{(1)}_3$. This
combination occurs $\binom{4}{1,1,1,1}$ times in \RF{eq:4}, which
means, also recalling that the multiplicity of the connection is
$2^4$, that we obtain for the triplet interaction,
\begin{align} 
  \beta \FF (\text{\small Fig.~\ref{fig:4}(b)}) =& -\frac{1} {4!}
  \binom{4}{1,1,1,1} \frac{ C_1^{(1)} C_2^{(2)} C_2^{(3)} C_3^{(1)}}
        {(2 \sigma)^4} \nonumber \\ &\times 2^4 G^{12}_{ijk}
        G^{23}_{jkl} G^{32}_{lmnp} G^{21}_{mnpi} \ .
\end{align}
After index contraction and substituting the polarizabilities, we have
\begin{equation}
  \beta \FF (\text{\small Fig.~\ref{fig:4}(b)}) = +12 \cos \varphi^{\,2}_{13}
  \frac {R_1^2 R_2^{10} R_3^2} {r_{12}^{7}r_{23}^{7}}
  \ , \label{eq:trip4}
\end{equation}
where $\varphi^{\,2}_{13}$ is the angle between $\bfr_1 -\bfr_2$ and
$\bfr_2 -\bfr_3$, or in this case, the exterior angle of the triangle
formed by the particles at $\bfr_2$.  The quadruplet interaction of
Fig.~\ref{fig:4}(c) is similarly found as
\begin{align}
  \beta \FF (\text{\small Fig.~\ref{fig:4}(c)}) =& -\frac{1} {4!}
  \binom{4}{1,1,1,1} \frac{ C_1^{(1)} C_2^{(2)} C_3^{(3)} C_4^{(1)}}
        {(2 \sigma)^4} \nonumber \\ & \times 2^4 G^{12}_{ijk}
        G^{24}_{jkl} G^{43}_{lmnp} G^{31}_{mnpi} \label{eq:quad4}
        \\ =& -12 \cos (3\varphi^{\,2}_{41} +4\varphi^{\,3}_{14})
        \frac{ R_1^2 R_2^4 R_3^6 R_4^2} {r_{12}^3 r_{24}^3 r_{43}^4
          r_{31}^4 } \ , \nonumber
\end{align}
with an analogous definition of the angles.

In this section we have tried to illustrate the {\em Feynman rules} of
our expansion by specific examples. It should be clear that computing
interactions is a straightforward exercise in counting powers and
diagrams relevant for the desired order in the inverse separation. The
Green function products are easy to evaluate using complex coordinates
as discussed in Appendix~\ref{sec:complex} and they turn out to be
proportional to a product of powers of the distances and the cosine of
a combination of angles associated with the geometry of the particle
configuration, determined by the multipole orders involved.

\begin{figure}[tb]
  \includegraphics[scale=.7]{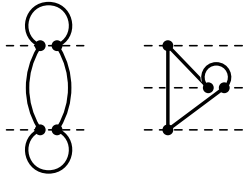} \caption{ Example diagrams with
    unphysical divergent self-energies.} \label{fig:self}
\end{figure}

The reader may have realized that there seem to be divergent pair and
triplet interactions in \RF{eq:4} due to one or more {\em self-links}
in the form $\del^n G(0)$ (\cf\ Fig.~\ref{fig:self}). These
divergences are all power-like ($\sim r^{-n}$ as $r \rightarrow 0$
where $n \ge 2$) and carry no physical information, as there is no
non-trivial renormalization group (RG) flow. The rigorous way to deal
with them is to add counter-terms to the Hamiltonian, aiming to cancel
the unphysical divergences from the free energy expansion. However,
since one can find all these counter-terms by renormalization
techniques, and the sole purpose of them is to remove the divergences
from the free energy expansion, the upshot of this systematic RG
treatment is that the divergences can simply be dismissed. We have
included Appendix~\ref{sec:RG} where the removal of one such
divergence by a counter-term is explicitly illustrated.

\subsubsection{Pair interactions} \label{sec:pair}

Although every multibody order is included in the cumulant expansion
(\ref{eq:cumex}), a systematic enumeration of all relevant diagrams
lies beyond the scope of this article. However, if only two particles
are involved, the solution can be worked out without too much trouble
and one can write down the complete asymptotic expansion of the free
energy.

Let us first establish that only even-numbered cumulants are relevant
for pair interactions. The reason is that there is no {\em physical}
(\ie\ finite) pair term in odd-numbered cumulants. The easiest way to
see this is to consider a pair diagram, \ie\ one with connections
between two world-lines such as Fig.~\ref{fig:4}(a), with an {\em odd}
number of vertices distributed along them: There is no possible
connection (Wick contraction) without at least one self-link. As we
discussed right before this subsection, such contributions to the free
energy are unphysical artifacts which can be removed
rigorously. Hence, pair interactions exclusively stem from
even-numbered cumulants.

\begin{figure}[b]
  \includegraphics[scale=.75]{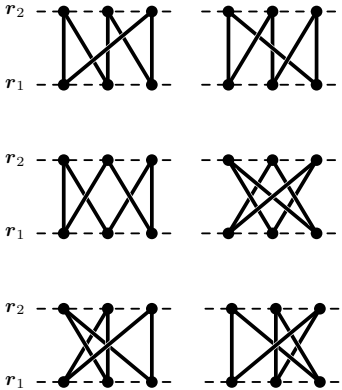} \caption{ Different
    ``stitches'' of the sixth cumulant pair
    interaction.} \label{fig:stitches}
\end{figure}

To write the full expansion of the pair free energy, one must be able
to do two things at an arbitrary cumulant order $2s$: $(i)$ enumerate
all contributing pair diagrams with $2s$ vertices and $(ii)$ evaluate
the diagrams, which involves doing the product of $2s$ propagators
[such as in Eqs.~(\ref{eq:pair4}), (\ref{eq:trip4}) or
(\ref{eq:quad4})] with any collection of multipole orders at the
vertices. Luckily, both of these can be done. The propagator product
is calculated in Appendix \ref{sec:complex}. As for the enumeration of
relevant diagrams, it is most suitable to again illustrate it with an
example: The sixth cumulant contains the diagrams depicted in
Fig.~\ref{fig:stitches}. These are all the connected pair diagrams
with six vertices, free of unphysical self-energies. As one can see by
starting from any one of the vertices and ``stitching'' the vertices
in all possible ways, there are $3! 2! /2 =6$ of these diagrams, which
generalizes to $g_s \mathrel {\mathop:} = s! (s-1)! /2$ for the
$(2s)^{\rm th}$ cumulant (except that $g_1 = 1$). With each given
collection of multipole orders at the vertices, the contribution of
each diagram to the cumulant, or the propagator product, will be
different (not in terms of power of $r$ but its prefactor) since the
partial derivatives will generally be distributed differently on the
propagators. While this might seem to complicate things, remember that
there is a sum over all possible collections of multipole orders out
front [\cf\ \RF{eq:4}]. Thus, the effect of these $g_s$ different
topologies is to repeat each scattering process $g_s$ times in the
sum. Hence one can consider only one of the $g_s$ possible diagrams
and multiply the result by $g_s$.

\begin{figure}[!tb]
  \includegraphics[scale=0.72]{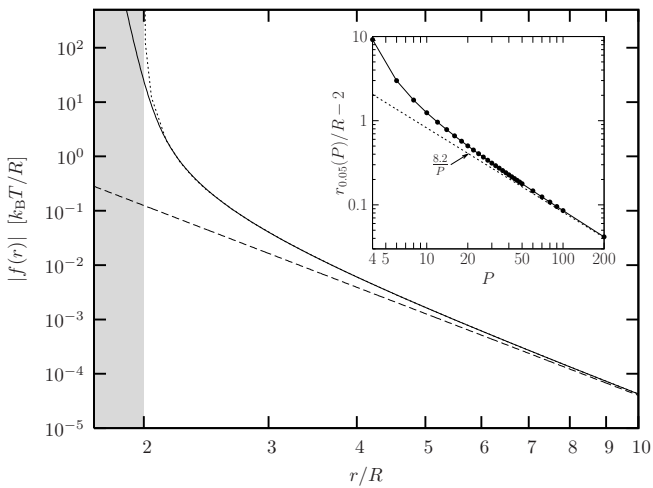} \caption{Comparison
    to numerical results of Ref.~\cite{LehOetPRE} on the attractive
    force $f(r) = - \del \FF / \del r$. The dotted and solid curves
    correspond respectively to Ref.~\cite{LehOetPRE} and our results
    (truncated at order $r^{-70}$ in the free energy). The dashed line
    represents the lowest order asymptotic solution, \ie\ the
    dipole-dipole interaction. The inset depicts how the power $P$ of
    the highest order term included in the expansion and the distance
    at which a 5 percent accuracy is achieved depend on each
    other.} \label{fig:Martin} 
\end{figure}

Taking everything together, the entire pair interaction can be written
as
\begin{align} 
  \beta \FF = -\sum_{\{\ell_i\}} & \left[ \frac{1} {2!} \binom{2}{1}
    g_1 \raisebox{-2.9ex} { \includegraphics[scale=.5] {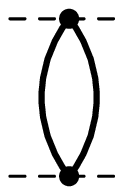}} +
    \frac{1} {4!}  \binom{4}{2} g_2 \raisebox{-2.9ex}
         {\includegraphics[scale=.5] {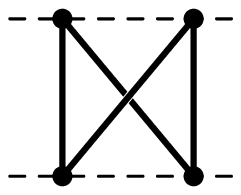}} \right. \nonumber
         \\ & \left. + \frac{1} {6!} \binom{6}{3} g_3
         \raisebox{-2.9ex} {\includegraphics[scale=.5] {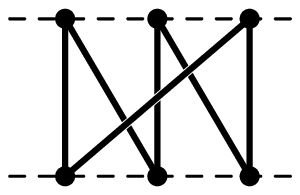}} +
         \ldots \right] \ , \label{eq:pairdiags}
\end{align}
where $\sum_{\{\ell_i\}}$ denotes a sum over all possible collections
of multipoles at the vertices.  Using the exact expressions for the
polarizabilities, \RF{eq:C}, and propagator products,
\RF{eq:GprodFull}, to evaluate the diagrams, the pair interaction can
be expressed in the following asymptotic expansion:
\begin{equation}
  \beta \FF = -\sum_{s=1}^{\infty} \frac{1}{s} \sum_{\ell_1, \ldots
    \ell_{2s}} \prod_{i=1}^{2s} \frac{ (\ell_i+\ell_{i+1} -1)!}
        {\ell_{i+1}! (\ell_i-1)!} \frac{R_1^{2L_{\rm o}} R_2^{2L_{\rm
              e}} } {r^{2(L_{\rm o} + L_{\rm e})}}
        \ , \label{eq:fullasym}
\end{equation}
where $\ell_{2s+1}=\ell_1$ , $ L_{\rm o} = \sum_{ {\rm odd\ } i}
\ell_i$ and $L_{\rm e} = \sum_{ {\rm even\ } i} \ell_i\ $. Assuming
identical particle radii to reduce clutter, the first few terms of
this expansion are
\begin{align} 
  \beta \FF =& - \left(\frac{R}{r} \right)^4 - 4 \left(\frac{R}{r}
  \right)^6 - \frac{31}{2} \left(\frac{R}{r} \right)^8 \nonumber
  \\ &-60 \left(\frac{R}{r} \right)^{10} - \frac{697}{3}
  \left(\frac{R}{r} \right)^{12} + \ldots \ . \label{eq:fewterms}
\end{align} 
Note that obtaining these numbers does not involve anything more
complicated than elementary algebra. The interaction up to order
$r^{-70}$ has been plotted in Fig.~\ref{fig:Martin}. A cursory
observation of the plot reveals that for (center-to-center) distances
larger than $r\approx4$, the lowest order term in \RF{eq:fewterms}
suffices. However, for closer separations, the number of higher order
terms one should include increases rapidly. We demonstrate this in the
inset to Fig.~\ref{fig:Martin} where $r_{5\%} (P)$ is defined as the
separation where the series (\ref{eq:fewterms}) terminated at $\OO
(r^{-P})$ achieves an accuracy of 5 percent. As the dimensionless
distance $d=r/R-2$ between the edges of the particles is decreased,
the power where the series can be truncated and still achieve less
than 5 percent error appears to asymptotically obey $P\approx 8.2\,
d^{-1}$. From this we see, for instance, that if the
surface-to-surface separation is a tenth of the particle radius,
\ie\ $d=0.1$, then one must retain all terms up to $\OO(r^{-82})$ in
order to achieve 5 percent accuracy.

\begin{figure}[!tb]
  \includegraphics[scale=.75] {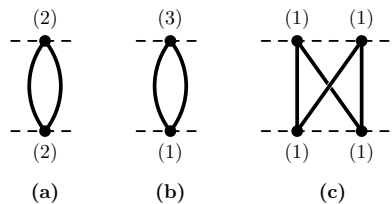} \caption{ (a) The
    quadrupole-quadrupole interaction $-9 (R/r)^{-8}$. (b) The
    dipole-octupole interaction $-6 (R/r)^{-8}$. (c) The
    dipole-dipole-dipole-dipole interaction $-(1/2)
    (R/r)^{-8}$.} \label{fig:other8} \vspace{-1ex}
\end{figure}

The first two terms of \RF{eq:fewterms} are the dipole-dipole and
dipole-quadrupole interactions also computed earlier by Lehle and
Oettel~\cite{LehOetPRE}. The same publication also contains a
quadrupole-quadrupole interaction of order $r^{-8}$, for the case of
tiltable colloids, of magnitude $-9$. Since the expansion
(\ref{eq:fewterms}) was developed under the assumption of frozen tilt
degree of freedom, the magnitude of the $r^{-8}$ term includes
additional interactions and is therefore different. Namely, these
additional interactions are the quadrupole-octupole and
dipole-dipole-dipole-dipole interactions (\cf\ Fig.~\ref{fig:other8})
with respective magnitudes $-6$ and $-1/2$, such that the resulting
total magnitude of the $r^{-8}$ term is $-9-6-1/2=-31/2$. Had we
instead assumed the colloids to be free to tilt, their dipole
polarizability $C^{(1)}$ would vanish---hence so would diagrams
Fig.~\ref{fig:other8}(b) and \ref{fig:other8}(c)---and the expansion
(\ref{eq:fewterms}) would have started with the quadrupole-quadrupole
term $-9(R/r)^8$.

Lehle and Oettel also compute a numerical solution to the interaction
for intermediate and close distances, where the first few terms of the
asymptotic expansion do not suffice~\cite{LehOetPRE}. As a
confirmation of our result, we show in Fig.~\ref{fig:Martin} the
interaction found in Ref.~\cite{LehOetPRE} by performing the
functional integrals numerically, with our expansion
(\ref{eq:fullasym}) up to order $r^{-70}$.  The agreement is
excellent, except upon approaching the contact distance $r=2R$, where
our truncated series remains finite while the numerical solution
appears to grow without limit. The exact solution indeed diverges as
\begin{equation}
  -\beta \FF = \frac {\pi^2} {24 \sqrt{d}} + \frac{1}{4} \log \left(
  \frac {4 d} {\pi^2} \right) - \frac {96-\pi^2} {576} \sqrt{d} +
  \OO(d) \ ,
\label{eq:close-divergence}
\end{equation}
where $d=(r-2R)/R$ is a scaled surface-to-surface distance. The
dominant term has previously been predicted by combining the exact
result for two parallel lines with the Derjaguin
approximation~\cite{Derjaguin,LehOetPRE}; the subleading contributions
cannot be extracted in this way but follow by expanding the exact
analytical solution which one of us recently found by using conformal
field theory techniques~\cite{bosonic}.

Since our formalism by construction produces an asymptotic expansion
in the inverse separation, it is clearly not the preferred method when
it comes to near-contact separations (see also inset in
Fig.~\ref{fig:Martin}). However, it turns out that even our large
distance expansion~(\ref{eq:fullasym}) is in a sufficiently manageable
form to permit extraction of the leading order of
(\ref{eq:close-divergence})---including the prefactor---by some
further analytical manipulation. We illustrate this in
Appendix~\ref{sec:PFA}.  This of course constitutes an application of
our expansion outside its region of highest analytical convenience,
but it vividly illustrates that ($i$) our expansion can indeed be
driven up to arbitrarily high order and that ($ii$) the resulting
predictions are correct up to the point where the exact solution
diverges.

\subsubsection{Beyond the pair interaction} \label{sec:beyond} 

When there are more than two particles involved, the free energy of
interaction generally does not decompose into a sum of pair
interactions but contains multibody interactions like the ones
discussed in Sec.~\ref{sec:Casimir}. As illustrated earlier, simple
power counting reveals to what order in the separations a certain
multi-body interaction might contribute, and evaluating the
interaction poses little difficulty.

Let us now discuss the leading order triplet interaction. A
dipole-dipole-dipole term from the third cumulant is the lowest order
triplet interaction conceivable. It is easy to see by power counting
that this interaction would scale as $r_{12}^{-2} r_{23}^{-2}
r_{31}^{-2}$. However, the strength of this interaction turns out to
be {\em identically zero}. This is due to the harmonicity of the Green
function of the problem, and applies to {\em any} diagram in an
odd-numbered cumulant (please see Appendix~\ref{sec:complex}).
Therefore, with this choice of surface Hamiltonian, many interactions
that would seem to exist based on scaling arguments actually
disappear~\footnote{This is not the case for a membrane, where the
  surface energy stems from bending elasticity~\cite{YolDesMemb}.}.

\begin{figure}[tb]
  \includegraphics[scale=.75] {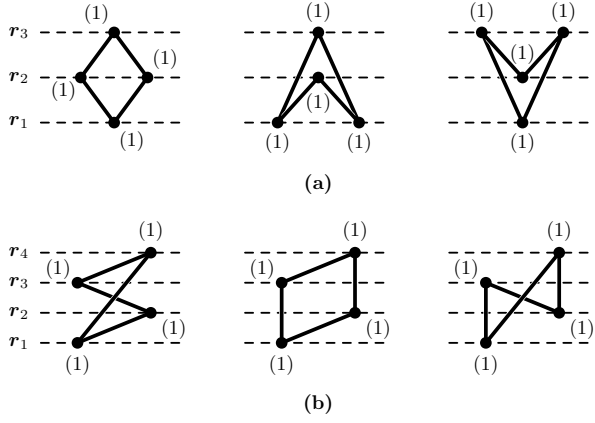} \caption{(a) The fourth
    cumulant diagrams contributing to the leading triplet
    interaction. (b) The fourth cumulant diagrams contributing to the
    leading quadruplet interaction.} \label{fig:tripquad}
\end{figure}

Since no triplet interaction exists in the third cumulant, one must
look for the leading triplet term in the next. The lowest order
interaction is between induced dipoles and stems from the diagrams
depicted in Fig.~\ref{fig:tripquad}(a). We have already computed a
subleading correction to this as an example in an earlier section,
that involved higher order multipoles. The leading order is similarly
computed and found to be
\begin{align}
  \beta \FF^{\rm trip}_{\rm lead} =& -\frac{1} {4!}  \binom{4}{1,2,1}
  \frac {C_1^{(1)} {C_2^{(1)}}^2 C_3^{(1)} } { (2\sigma)^4} 2^4
  G^{12}_{ij} G^{23}_{jk} G^{32}_{kl} G^{21}_{li} \nonumber \\ &+
  \text{\small cyc. perm. of $(1 2 3)$}
\end{align}
which becomes, after evaluating the propagator product---either by
direct computation or using \RF{eq:GprodFull}---and substituting the
polarizabilities from \RF{eq:C},
\begin{equation}
  \beta \FF^{\rm trip}_{\rm lead} = - \frac{ R_1^2 R_2^4 R_3^2}
        {r_{12}^4 r_{23}^4} - \frac{ R_1^2 R_2^2 R_3^4} {r_{23}^4
          r_{31}^4} - \frac{ R_1^4 R_2^2 R_3^2} {r_{31}^4 r_{12}^4}
        \ .
\end{equation}

The fourth cumulant is also where the leading order quadruplet
interaction resides. This results from the diagrams shown in
Fig.~\ref{fig:tripquad}(b) among four induced dipoles. It follows as,
\begin{widetext}
\begin{align}
  \beta \FF^{\rm quad}_{\rm lead} = & -\frac{1}{4!} \binom{4}{1,1,1,1}
  \frac {C^{(1)}_1 C^{(1)}_2 C^{(1)}_3 C^{(1)}_4} { (2 \sigma)^4} 2^4
  \left( G^{12}_{ij} G^{23}_{jk} G^{34}_{kl} G^{41}_{li} + G^{12}_{ij}
  G^{24}_{jk} G^{43}_{kl} G^{31}_{li} +G^{13}_{ij} G^{32}_{jk}
  G^{24}_{kl} G^{41}_{li} \right) \\ =& -2 R_1^2 R_2^2 R_3^2 R_4^2
  \left[ \frac{ \cos ( \varphi^{\,2}_{13} - \varphi^{\,3}_{24} +
      \varphi^{\,4}_{31} - \varphi^{\,1}_{42}) } {r_{12}^2 r_{23}^2
      r_{34}^2 r_{41}^2} + \frac{ \cos ( \varphi^{\,2}_{14} -
      \varphi^{\,4}_{23} + \varphi^{\,3}_{41} - \varphi^{\,1}_{32}) }
    {r_{24}^2 r_{43}^2 r_{31}^2 r_{12}^2} + \frac{ \cos (
      \varphi^{\,3}_{12} - \varphi^{\,2}_{34} + \varphi^{\,4}_{21} -
      \varphi^{\,1}_{43}) } {r_{32}^2 r_{24}^2 r_{41}^2 r_{13}^2}
    \right] \ . \nonumber
\end{align}
\end{widetext}

Depending on the desired level of accuracy, corrections to these
interactions as well as further multiplets may be required. In this
case, one simply identifies the relevant diagrams or interactions by
power-counting and the computation that follows is fairly
straightforward as shown.

\subsection{On anisotropic and flexible objects}

We have carried out our discussion under the assumption of rigid and
circular particles. An obvious extension is towards particles with
internal flexibility or non-axisymmetric shapes. Although a thorough
treatment of these extensions can be quite involved, there are
nevertheless certain simple but rigorous remarks we can make at this
stage.

Not much changes when the particles considered are flexible (or the
constraints on their motion are not completely rigid). Since the
boundary conditions at the circumference of the particles change
accordingly, fixing the polarizabilities will require the solution of
a different boundary value problem on the full theory side of the
matching procedure, but the form of the effective Hamiltonian is
exactly the same. The subsequent change in the values of the
polarizabilities will result in different interaction
strengths. However, this does not affect one's ability to identify
features of the interaction, such as what the leading power of inverse
separations in a given multi-body interaction is.

If the elastic properties of the particles are ever so slightly
different from the surface, then the free energy expansion acquires
another smallness parameter intrinsic to the polarizabilities. This
renders high cumulants in $\beta\FF$ negligible compared to the
second; a so-called weak coupling~\cite{MilWagPRD} approximation.  In
such a case, being able to obtain a closed form result is more likely
since only the second cumulant has to be taken into account. This is
one of the limits considered in Ref.~\cite{LinZanMohPryPRL}. These
authors restrict to two scatterings, analogously to truncating the
cumulant expansion after the second, which is justified by the
weakness of the perturbations.

In the case of an anisotropic particle, the response of the particle
to incident backgrounds depends on its orientation, which is clearly
not the case for the response given in \RF{eq:response}. One now needs
to consider polarizability {\em tensors} to describe the anisotropic
response of a non-circular particle. The first few terms in the
derivative expansion of a particle read
\begin{equation}
  \Delta \HH_a = \half \left. \left( \ff_i C_{ij}^{(1)} \ff_j +
  \ff_{ij} C_{ijkl}^{(2)} \ff_{kl} + \ldots \right) \right
  \vert_{\bfr=\bfr_a} \ .
\end{equation}
Fixing the polarizability tensors will require solving more
complicated boundary value problems, but the lowest order interaction
will still stem from a pair of induced dipoles (tensor $C^{(1)}_{ij}$)
and therefore scale as $r^{-4}$, no matter what the shape of the
particle might be. Or if the particle is free to tilt, then the
polarizability of any term involving an $\ff_i$ will vanish, thereby
making the leading asymptotic interaction a quadrupole-quadrupole at
order $r^{-8}$. These are indeed in line with the findings of
Ref.~\cite{NorOetPRE}, where explicit calculations of the leading
order interactions between rigid ellipsoidal particles were presented.

\section{On corn flakes}

So far, we have focused on interactions initiated by thermal
fluctuations and have thus restricted to particles that do not deform
the film in its ground state. However, it is straightforward to
include such effects in our formalism.  This kind of situation
naturally arises from an irregular three phase contact line between a
colloid and the fluid-fluid interface it is trapped
at~\cite{StaDusJohPRE}. Macroscopically, we see corn flakes perturbing
the surface of a bowl of milk in a similar way. Again, one would
expect (and can indeed observe) that once these deformations overlap,
they give rise to aggregation of the flakes.

These perturbations are treated as ``permanent charges'' in the
effective theory, beside the induced charges arising from
constraints. The effective theory incurs local linear terms in
addition to the quadratic ones upon this modification. The easiest way
to determine these linear terms is again through an argument involving
the response of each particle to a background. Consider a particle
whose contact line height profile has a nontrivial multipole $\ff^{\rm
  cl} (\varphi) = \eta_\ell \cos \left( \ell \varphi - \ell
\alpha_\ell \right)$ around its center with a phase angle
$\alpha_\ell$. When a background $ h^{\rm bg} (\bfr) = \eta_\ell
R^{-\ell} r^\ell \cos \left( \ell \varphi - \ell \alpha_\ell \right)$
is incident on this particle, the boundary condition $\left. (\ff^{\rm
  in} = \ff^{\rm out}) \right|_{r=R}$ is automatically satisfied by
the fact that $\ff^{\rm bg} (R, \varphi) = \ff^{\rm cl} (\varphi)$ and
hence {\em no} net response will be triggered. Another way to phrase
this is that the field of the permanent charge and the induced
response to this specific background eliminate each other. Therefore,
if this ``preferred background'' is denoted by $p^{(\ell)}(\bfr)$, the
local linear terms representing the permanent charges can be encoded
into the quadratic induced charge terms as a shift:
\begin{equation}
  \Delta \HH [\ff] = \half \sum_{a,\ell} C^{(\ell)}_a \left[ \del^\ell
    \ff (\bfr_a)- \del^\ell p^{(\ell)} (\bfr_a) \right]^2 \ .
  \label{eq:DeltaHflakes}
\end{equation}
Note that the values of the polarizability coefficients are the same,
whether the preferred shape of the particles are flat or curved, as
long as the constraints are the same. Notice that in the absence of
(vertical) external forces, the vertical movement of a particle is
free and hence $C^{(0)}=0$. Similarly, in the absence of (horizontal)
external torques, the tilt motion is free and $C^{(1)}=0$. In the
following, we will assume that this is the case and therefore the
multipole expansion starts at $\ell=2$.

From \RF{eq:DeltaHflakes} one sees that the linear terms in the
effective Hamiltonian are
\begin{equation}
  -\sum_{a,\ell} C_a^{(\ell)} \del^\ell p^{(\ell)} (\bfr_a)
  \del^{\ell} \ff (\bfr_a) \ ,
\end{equation}
which encode the permanent sources
\begin{equation}
  -\sum_{a,\ell} (-)^\ell C^{(\ell)}_a \del^\ell \left[ \delta (\bfr
    -\bfr_a) \, \del^\ell p^{(\ell)} (\bfr_a) \right] \ .
\end{equation}
In the diagrammatic expansion, the linear terms correspond to vertices
with only one link attached. Therefore, interactions involving
permanent charges are {\em open} diagrams initiating and terminating
at these. Such diagrams have one link less but the same number of
vertices compared to the fluctuation-induced interactions, and power
counting shows the free energy due to them will scale as $\FF \sim
\sigma R^2$, \ie\ with no temperature dependence. These {\em ground
  state} contributions together with the fluctuation part, $\sim \kB T
$, of the previous section result in a free energy of the form $\FF =
a_0 + a_1 T$, which allows us to view this separation into ground
state energy and fluctuation correction as equivalent to a separation
into the energy and entropy terms in $\FF= E-TS$. Observe that this
equivalence would not hold had we considered anharmonicities in the
theory and thus furnished the free energy with contributions higher
than linear order in the temperature.

For the sake of clarity, let us consider a pair of particles with a
saddle shaped contact line deformation. That is, each particle will
have a preferred background of the form $p(\bfr) = (1/2) k r^2 \cos
(2\varphi - 2\alpha)$ around its center, where the angle $\alpha$
describes the orientation of the saddle. For a colloid under no
external vertical force (\eg\ gravity) or horizontal torque, this
quadrupole deformation is the lowest order multipole contact line
irregularity that can occur. The linear part of the effective theory
follows as $-\sum_{a} C^{(2)}_a p_{ij} (\bfr_a) \ff_{ij} (\bfr_a)$,
where
\begin{equation}
  p_{ij} (\bfr_a) = k_a \left[ \begin{array} {lr} \cos 2\alpha_a &
      \sin 2\alpha_a \\ \sin 2\alpha_a & -\cos 2\alpha_a \end{array}
    \right] \ .
\end{equation}
Notice the simple geometric interpretation of $k_a$ as the magnitudes
of each principal curvature of the saddles.

\begin{figure}[!tb]
  \includegraphics[scale=.75] {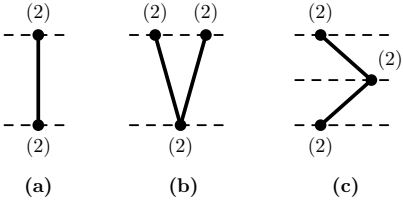} \caption{(a) The direct
    interaction $ E^{(4)} \sim r^{-4}$ between two permanent
    quadrupoles. (b) The pair interaction $ E^{(8)} \sim r^{-8}$
    between a permanent quadrupole and the quadrupole it induces on
    the other particle. This is the lowest order energetic pair
    interaction that involves scattering from an induced charge. (c)
    At the same order as diagram (b), there also exists a triplet
    interaction $E^{(8)}_{\text{trip}}$involving two separate
    permanent charges and one induced charge. } \label{fig:staple}
\end{figure}

The leading interaction stems from the second cumulant as a direct
interaction between the two permanent quadrupoles,
Fig.~\ref{fig:staple}(a), easily found as
\begin{align}
  E^{(4)} = &-\frac{1}{2!} \binom{2}{1,1} \frac{ C_1^{(2)} C_2^{(2)}}
  {\sigma} p_{ij} (\bfr_1) G^{12}_{ijkl} \, p_{kl} (\bfr_2) \nonumber
  \\ =& -3 \pi \sigma \cos (2 \alpha_1 + 2\alpha_2) \frac{ k_1 R_1^4
    R_2^4 \, k_2} {r^4} \ , \label{eq:Stamou}
\end{align}
where the pair was assumed to lie on the $x$ axis to declutter the
expression. The functional dependence of the interaction implies both
an attractive force and vertical torques on the saddle-like
particles. The same potential was found by Stamou {\em et
  al.}~\cite{StaDusJohPRE}. We have seen earlier that, between
colloids that are free to tilt and fluctuate vertically, the leading
fluctuation-induced interaction is of order $r^{-8}$, which can only
compete with the interaction (\ref{eq:Stamou}) at close separations.

Although we have restricted to a toy problem where the only permanent
deformation is quadrupolar, it should be obvious how one would handle
higher order multipoles of the contact line irregularity for accuracy
at closer separations. However, one should note that the true
interaction potential is not merely a superposition of direct
interactions between permanent charges, similar to
Fig.~\ref{fig:staple}(a), but includes scatterings from induced
charges, such as Fig.~\ref{fig:staple}(b). The former is equivalent to
assuming the total shape of the surface is given by a superposition of
the deformations caused by each particle independently of every other,
\ie\ neglecting their boundary conditions, which is why it misses
induced charges. Such a superposition approximation is captured by the
second cumulant and holds up to the order the next cumulant begins to
contribute. This is the order $r^{-8}$, where the third cumulant
contributes the diagram in Fig.~\ref{fig:staple}(b), evaluating to the
repulsion
\begin{equation}
  E^{(8)} =\frac{9}{2} \pi \sigma (k_1^2 R_1^4 + k_2^2 R_2^4)
  \frac{R_1^4 R_2^4} {r^8} \ ,
\end{equation}
which is the lowest order pair interaction that involves a scattering
from an induced charge. At the same order, one observes the first
multibody interaction as well, which is depicted in
Fig.~\ref{fig:staple}(c), and evaluates to $E^{(8)}_{\text{trip}} =
E^{(8)}_{123} + E^{(8)}_{231} + E^{(8)}_{312}$ where
\begin{align}
  E^{(8)}_{123} =9 \pi \sigma \frac{ k_1 R_1^4 R_2^4 R_3^4 \, k_3}
  {r_{12}^4 r_{32}^4} \cos ( 4\varphi^{\,2}_{13} - 2\alpha_1
  +2\alpha_2 )
\end{align}
and so on. If the particles possess higher multipole order contact
line irregularities, merely taking these into account within a
superposition approximation does not capture every possible correction
systematically, since it misses contributions such as
Figs.~\ref{fig:staple}(b) and \ref{fig:staple}(c) by construction.

To summarize, although we primarily use the EFT formalism to
facilitate the evaluation of a partition function, the ground state
interactions mediated by the surface can easily be calculated as
well. One can readily encode the permanent deformations imposed by the
colloids as linear terms in the effective theory, upon which the
ground state energy strictly separates from the fluctuation
correction. Computations of this energy that rely upon the
superposition of perturbations caused by the particles only hold for
direct interactions of the second cumulant.

\section{Conclusion}

We have presented and discussed a formalism to compute the
surface-mediated forces between rigid disks trapped at an interface,
based on effective field theory. This formalism reduces each finite
sized particle to a point, which greatly simplifies
computations. Still, these points are by explicit construction
equipped with all the information necessary to capture their finite
sized counterpart's behavior, and hence the treatment is not
approximate. In particular, their size is not simply recovered as the
continuum theory's ad hoc ultraviolet cutoff---an approach which
can neither conceptually satisfy nor be plausibly extended to more
than one particle radii.

The extension from the free surface theory to the effective theory is
achieved by the addition of a derivative expansion $\Delta \HH$, with
extra terms localized at the centers of the particles, encoding the
constraints they enforce on the fluctuations as induced capillary
multipoles. Symmetries determine the form of this extension to the
Hamiltonian, while one still has to fix the values of undetermined
polarizability coefficients via a matching procedure. However, one
gains insight into the interaction free energy even without fixing
these coefficients, such as the leading order distance dependence of
the force or the vanishing of certain many-body interactions due to
the kernel of the theory. In this article we focused on circularly
symmetric particles, which allowed us to determine the effective
Hamiltonian completely, \ie\ to all derivative orders. However, in
cases where this is not possible and a certain level of accuracy ---
in powers of $r$---is desired, knowing only a fraction of all the
polarizability coefficients will suffice.

After the effective theory has been established, the interaction free
energy is then computed in a cumulant expansion. $N$-body interactions
are identified as Feynman diagrams linking $N$ world-lines and are
computed in a fairly straightforward manner. The two expansions, the
cumulant expansion and that in the local derivatives of the field at
the inclusions, afford a transparent interpretation of every term that
contributes to the interaction as a scattering process between various
multipole moments of the induced charges, similar in spirit to
Ref.~\cite{EmiGraJafKarPRD}. Unless both of these expansions are
performed, consistent results cannot be achieved. Also, owing to
simplifications due to the harmonicity of the theory, the number of
dimensions, and the symmetry of the particles, a general expression
for any possible diagram was obtained in Appendix
\ref{sec:complex}. We have used this expression to develop a full
asymptotic expansion of the pair interaction. It was possible to
partially sum up the expansion. We believe a complete summation of the
series to obtain a closed form expression may be possible as well,
though we have not succeeded here.

As we have discussed, an extension to non-rigid and non-circular
objects entails no major change in the formalism. The former only
affects the strength of the polarizability coefficients and therefore
the interactions, while the latter requires consideration of tensor
polarizabilities, and both fixing their values and computing
interactions become a bit more involved.

Moreover, we have presented results on ground state interactions
between the particles as well as the fluctuation-induced
interactions. This required us to add permanent charges to the
effective theory, which is simply achieved by shifting the zero of the
induced charge terms. The ground state interactions then strictly
separate from the fluctuation corrections and can be computed in the
same manner we discussed for the fluctuation-induced forces. In
addition to deriving the lowest order pair interaction, in agreement
with known results, we illustrated how corrections---higher order or
multibody---are computed.

Note that the same formalism can be used to treat surfaces of
different characters, \ie\ different free surface Hamiltonians. A
relevant example is the problem of surface-mediated interactions
between inclusions on a biological membrane, which we will treat in a
future publication~\cite{YolDesMemb}. Additionally, fluctuation
corrections of higher-than-linear order in $\kB T$ can be
systematically studied by increasing the anharmonicity of the theory,
\eg\ relaxing the weak deformation assumption on the interface or the
constant footprint of the particles.

\begin{appendix}

\section{Index contractions in complex coordinates} \label{sec:complex}

In this appendix we will go into details of how to perform the index
contractions encountered in computing induced fields and propagator
products by transforming to complex coordinates, $z=x+\im y$ and
$\zb=x-\im y$. We first note that in cartesian coordinates, covariant
and contravariant elements of tensors are identical, \ie\ the metric
is the identity. Therefore we have not distinguished between them and
have written all indices as subscripts in the main body of the
article. However, now that we are considering a change of coordinates,
we will use a covariant notation instead. For example,
\RF{eq:response} will read
\begin{align}
  \delta \ff (\bfr) =& -\frac{ C^{(\ell)}} {\sigma} \ff^{\rm bg}_{i_1
    \ldots i_\ell} (0) G^{i_1 \ldots i_\ell} (0,\bfr) \nonumber \\ =&
  -\frac{ C^{(\ell)}} {\sigma} g^{i_1 j_1} \ldots g^{i_\ell j_\ell}
  \ff_{i_1 \ldots i_\ell}^{\rm bg} (0) G_{j_1 \ldots j_\ell} (0, \bfr)
  \ , \label{eq:contrxn}
\end{align}
where $g^{ij}$ is the (inverse) metric tensor.

From the embedding $x=(1/2)(z+\zb), y=(1/2\im) (z-\zb)$ the
tangent vectors $\vec{e}_z=(1/2)(1,-\im)$ and $\vec{e}_{\zb} =
(1/2)(1,\im)$ follow. Then the metric tensor $g_{ij} = \vec{e}_i
\cdot \vec{e}_j$ and its inverse are found as
\begin{equation}
  g_{ij} = \half \left( \begin{array} {cc} 0 &1\\1&0 \end{array}
  \right) \ , \ g^{ij} = 2 \left( \begin{array} {cc} 0
    &1\\1&0 \end{array} \right) \ . \label{eq:metric}
\end{equation}
Since the metric is constant, the Christoffel symbols vanish, and
therefore covariant derivatives are just partial derivatives (and
hence also commute). 

The reason why it is useful to go to complex coordinates is as
follows. The index contractions needed for our computations all
involve partial derivatives of the harmonic Green function, which is
\begin{equation}
  G({\boldsymbol z},{\boldsymbol z}') = -\frac{1}{4\pi} \log (z-z')
  (\zb -\zb') \ ,
\end{equation}
in complex coordinates. It is easily seen that
\begin{equation}
  \del_z \del_{\zb} G =0 \ . \label{eq:delzdelzb}
\end{equation}
which is equivalent to the harmonic property of the Green function
since $\nabla^2 G = g^{ij} G_{ij} = 4 \del_z \del_{\zb} G$. This is a
useful property because, when expressed in complex coordinates, all
index combinations involving an alternation of indices on $G$ vanish
from the contraction, leaving only terms that involve either
\begin{equation}
  \frac{ \del^n G} {\del z^n} = \frac{ (n-1)!} {4\pi} \frac{1}
       {(z'-z)^n} \ , \label{eq:delznG}
\end{equation}
or its complex conjugate. This property and the simple form of the
metric tensor (\ref{eq:metric}) simplify things greatly, allowing us
to express products involving an arbitrary number of indices and
factors.

The computation of the induced deformation, \RF{eq:response} or
\RF{eq:contrxn}, is now as follows: The background $\ff^{\rm bg} =
\alpha r^\ell \cos \ell \varphi$ can be rewritten in complex
coordinates as $\ff^{\rm bg} = (\alpha/2) (z^\ell + \zb^\ell)$. Using
Eqs.~(\ref{eq:metric}) and (\ref{eq:delzdelzb}) we find
\begin{align}
  \delta \ff (\bfr) =& -\frac{ C^{(\ell)}} {\sigma} \ff^{\rm bg}_{i_1
    \ldots i_\ell} (0) G^{i_1 \ldots i_\ell} (0,{\boldsymbol z})
  \nonumber \\ =& -\frac{ C^{(\ell)}} {\sigma} g^{i_1 j_1} \ldots
  g^{i_\ell j_\ell} \ff^{\rm bg}_{i_1 \ldots i_\ell} (0) G_{j_1 \ldots
    j_\ell} (0,{\boldsymbol z}) \nonumber \\ =& -\frac{ C^{(\ell)}}
  {\sigma} 2^\ell \del_{\zb}^\ell \ff^{\rm bg} (0) \del_z^\ell G
  (0,{\boldsymbol z}) + \text{c.\,c.} \label{eq:appeffresp1}
\end{align}
Note that the index contraction, which generally has $2^\ell$ terms,
was reduced to only two terms, owing to the property $\del_z
\del_{\zb} G =0$. Now, using \RF{eq:delznG} the result we used in
\RF{eq:effresp} can be found:
\begin{align}
  \delta \ff (\bfr) =& -\frac{ C^{(\ell)}} {\sigma} 2^\ell \frac{
    \alpha \ell!}  {2} \frac{(\ell-1)!} {4\pi} \left( \frac{1}{z^\ell}
  + \frac{1} {\zb^\ell} \right) \nonumber \\ =& -C^{(\ell)} \frac{
    2^\ell \ell! (\ell-1)! \alpha} {4\pi \sigma} \frac{ \cos \ell
    \varphi} {r^\ell} \ . \label{eq:appeffresp}
\end{align}
Note that one has to pay attention to which argument of the Green
function is differentiated so as to avoid sign errors.

The other point where we encounter index contractions is in propagator
products such as
\begin{equation} \label{eq:4prod}
  \del^{\ell_a} \del^{\ell_b} G^{ab} \del^{\ell_b} \del^{\ell_c}
  G^{bc} \del^{\ell_c} \del^{\ell_d} G^{cd} \del^{\ell_d}
  \del^{\ell_a} G^{da} \ ,
\end{equation}
where $\del^\ell$ is a shorthand for $\del_{i_1} \del_{i_2} \ldots
\del_{i_\ell}$. To make the underlying algebra more transparent, let
us first look at this product of four propagators, before we
generalize to an arbitrary number. Similarly to the derivation of
Eqs.~(\ref{eq:appeffresp1}) and (\ref{eq:appeffresp}), and defining
${\boldsymbol z}_{ab} = {\boldsymbol z}_a - {\boldsymbol z}_b$, this
product can be written as
\begin{align}
    \text{(\ref{eq:4prod})} =& 2^{\ell_a +\ell_b +\ell_c + \ell_d}
    \del^{\ell_a}_z \del^{\ell_b}_z G({\boldsymbol z}_{ab})
    \del^{\ell_b}_{\zb} \del^{\ell_c}_{\zb} G({\boldsymbol z}_{bc})
    \nonumber \\ & \times \del^{\ell_c}_z \del^{\ell_d}_z
    G({\boldsymbol z}_{cd}) \del^{\ell_d}_{\zb} \del^{\ell_a}_{\zb}
    G({\boldsymbol z}_{da}) +\text{c.\,c.} \label{eq:GFprodz}
\end{align}
Now we substitute from \RF{eq:delznG} to find
\begin{widetext}
\begin{align}
  \text{(\ref{eq:4prod})} =& \frac{ (\ell_a+\ell_b-1)!} {4\pi}
  \frac{(\ell_b+\ell_c-1)!} {4\pi} \frac{ (\ell_c+\ell_d-1)!} {4\pi}
  \frac{(\ell_d+\ell_a-1)!} {4\pi} \frac{ (-2)^{\ell_a +\ell_b+
      \ell_c +\ell_d} } {z_{ab}^{\ell_a+\ell_b}
    \zb_{bc}^{\ell_b+\ell_c} z_{cd}^{\ell_c+\ell_d}
    \zb_{da}^{\ell_d+\ell_a}} + \text{c.\,c.} \nonumber \\ =& \frac{
    (\ell_a+\ell_b-1)! (\ell_b+\ell_c-1)! (\ell_c+\ell_d-1)!
    (\ell_d+\ell_a-1)!} {(4\pi)^4 2^{ -(\ell_a+ \ell_b +\ell_c
      +\ell_d+1)}} \frac{ \cos( \ell_b \varphi^{\,b}_{ac} - \ell_c
    \varphi^{\,c}_{bd} + \ell_d \varphi^{\,d}_{ca} - \ell_a
    \varphi^{\,a}_{db})} {r_{ab}^{\ell_a+\ell_b}
    r_{bc}^{\ell_b+\ell_c} r_{cd}^{\ell_c+\ell_d}
    r_{da}^{\ell_d+\ell_a}} \ ,
\end{align}
\end{widetext}
where $z_{ab} = \mathrel{\mathop:} r_{ab} \ee^{\im \varphi_{ab}}$ and
$\varphi^{\,b}_{ac} \mathrel{\mathop:} = \varphi_{ab} - \varphi_{bc}$
is the angle between $\bfr_a -\bfr_b$ and $\bfr_b - \bfr_c$, \etc. It
is obvious how this result generalizes to the product of an arbitrary
number $q$ of propagators:
\begin{equation} \label{eq:GprodFull} 
  2\, {\rm Re} \prod_{i=1}^{q} \frac{ (-2)^{\ell_i} (\ell_i + \ell_{i+1}
    -1)!} {4\pi } \frac{ \ee^{ \im (-)^i \ell_i \varphi^{\, a_i}_{
        a_{i-1}a_{i+1}}}} {r_{\!\!\!  \phantom{1} a_i a_{i+1}}^{\ell_i
      + \ell_{i+1}} }
\end{equation}
with $\ell_{q+1} = \ell_1$. We use this formula to evaluate many
interactions in Section~\ref{sec:Casimir}, most noticeably in
\RF{eq:fullasym} where a pair interaction is considered and thus the
angles are zero.

Lastly, we would like to revisit a statement we made in Section
\ref{sec:beyond}, namely that the free energy expansion
(\ref{eq:cumex}) for the surface tension Hamiltonian has identically
vanishing odd cumulants. This is easily seen to hold by observing that
it is not possible to write a {\em nonzero} expression analogous to
(\ref{eq:GFprodz}) for an odd number of propagators, due to the
property (\ref{eq:delzdelzb}) and the off-diagonal form of the
metric. As it rests on \RF{eq:delzdelzb}, or $\nabla^2 G =0$, this
property of all odd numbered cumulants vanishing is special to the
harmonic free surface Hamiltonian.

\section{Near-contact asymptotics} \label{sec:PFA}

In problems where the field equation, the boundary conditions
\etc.\ are more complicated than those considered in this article, one
generally does not expect to calculate the cumulant expansion in its
entirety. However, in this specific case, it is possible to obtain an
exact and complete expansion: Eq.~(\ref{eq:fullasym}). While finding a
closed form for the whole series or at least the expansion
coefficients is a difficult task, considerable simplifications turn
out to be possible.  The inner $2s$-fold sum of \RF{eq:fullasym} has
the form of a so-called \emph{binomial cycle}~\cite{RiordanCombId} and
it is possible to perform all but one of the sums adapting from
Ref.~\cite{RiordanCombId}.  Defining $u=R^2/r^2$ for the case of
identical radii, \RF{eq:fullasym} reduces to
\begin{equation}
  \beta \FF = -\sum_{s,n=0}^\infty \binom{2n}{n} \frac{u^{2s+n+2} \,
    \hyperF \!  \left(-n,1;-2n; \tfrac{f_{2s}(u)} {f_{2s+1}(u)}
    \right)} {(s+1) f_{2s+1}(u) f_{2s}(u) } \ , \label{eq:summed}
\end{equation}
where $\hyperF (a,b;c;x)$ is the hypergeometric
function~\cite{AbramStegun} and the polynomials $f_q(u)$ have the
recursion relation
\begin{equation}
  f_q (u) = f_{q-1} (u)-u f_{q-2}(u) \ ,
\end{equation}
with $f_0=f_{-1}=1$, related to continued fractions. Also note that
the summation index $s$ merely labels cumulants and is related to the
cumulant order $q$ as $q=2(s+1)$. This already amounts to a big
practical improvement over \RF{eq:fullasym}, because \RF{eq:summed}
can be \emph{much} more efficiently expanded in powers of inverse
distance.

From the recursion relation of $f_q$, one can also derive the expression
\begin{equation}
  f_q (u) = \sum_{i=0}^{\lfloor \frac{q+1}{2} \rfloor}
  \binom{q+1-i}{i} (-u)^i \ ,
\end{equation}
where $\lfloor \cdot \rfloor$ denotes the integer part, or better
still, recognize (in our case, {\footnotesize MATHEMATICA} did this
for us) its closed form as
\begin{equation}
  f_q (u) = \frac{\chi^{q+2} -1} {(\chi^2 -1) (\chi^2 +1 )^q}
  \ , \label{eq:fclosed}
\end{equation}
where $\chi(u) =( 1-\sqrt{1-4u} -2u)/2u$. Thanks to the closed-form
expression for $f_q$, one can show that in the interval $0<u<1/4$
(\ie\ from asympotically separated to osculating disks, recalling
$u=R^2/r^2$), the ratio $f_{2s} (u) / f_{2s+1} (u) < 2$, which is the
condition for the hypergeometric function in Eq.~(\ref{eq:summed}) to
converge as $n \rightarrow \infty$:
\begin{equation}
  \lim_{n \rightarrow \infty} \hyperF (-n,1;-2n;x) = \left( 1 -
  \frac{x}{2} \right)^{-1} \ . \label{eq:hyperlim}
\end{equation}
One can then use Eq.~(\ref{eq:hyperlim}) to isolate the leading
divergence in each cumulant $q=2(s+1)$ in Eq.~(\ref{eq:summed}) as the
$n \rightarrow \infty$ tail of the inner sum and obtain the resulting
divergence in $\beta \FF$ as
\begin{equation}
  \beta \FF_{\infty} \sim -\sum_{s=0}^{\infty} \frac{u^{2s+2}
    \sum_{n=0}^{\infty} \binom{2n}{n}u^n} {(s+1) f_{2s}(u) \left(
    f_{2s+1}(u) - \frac{f_{2s}(u)}{2} \right)} \ . \label{eq:Fdiv}
\end{equation}
The symbol ``$\sim$'' reminds us that this only captures the leading
order asymptotic divergence.

After noting that $\sum_{n=0}^{\infty} \binom{2n}{n} u^n =
(1-4u)^{-\half}$, we will define the (scaled) surface-to-surface
separation $d$ between the disks as $u=1/(d+2)^2$ and re-express
Eq.~(\ref{eq:Fdiv}) for $d \ll 1$ using Eq.~(\ref{eq:fclosed}) in
order to obtain the leading divergence in $\beta\FF_{\infty}$:
\begin{equation} \beta \FF_{\infty} \sim -\frac{1}{4 \sqrt{d}}
    \sum_{s=1}^{\infty} \frac{1}{s^2} = -\frac {\zeta(2)} {4 \sqrt{d}}
    =-\frac{\pi^2}{24 \sqrt{d}} \ , \label{eq:PFA}
\end{equation}
in agreement with \RF{eq:close-divergence}.

The crudeness of isolating the divergent part only allows one to
obtain the leading order proximity asymptotics. In a recent
preprint~\cite{bosonic} one of us used conformal field theory to
exactly compute the partition function of a bosonic field on a plane
with two holes. This is isomorphic to the problem of two disks on a
film studied in this article, and the interaction can in fact be
written in closed form (it involves a Dedekind $\eta$-function). Its
expansion at large distances reproduces our EFT series, while an
expansion at contact leads to \RF{eq:close-divergence}.

We again point out that computing the close distance divergence is not
the most obvious application for our specific EFT implementation
(which is built on a large distance expansion), but it nevertheless
illustrates the power of the formalism quite vividly.

\section{Cancellation of divergences by counter-terms} \label{sec:RG}

In this appendix we will illustrate how self-energy divergences can be
removed by counter-terms. While this is, to some extent, a standard
textbook affair, we include it in order to illustrate the mechanism in
the presently relevant context for readers not necessarily familiar
with (quantum) field theory.

First of all, let us clarify which divergences are of interest
here. Purely numerical divergences, such as those of the first
cumulant, get absorbed in the definition of the free surface energy
(\ie\ no particles); they do not refer to the relative positions of
the particles and thus do not affect interaction energies. Similar
trivial divergences occur in every cumulant order and do not require
any further discussion. The more interesting divergences are those
which multiply interactions between particles and thus cannot be
lumped together with the free surface energy.

The appearance of these divergences is due to the very construction of
an effective field theory. The self-interactions necessarily involve
the short distance (high energy) physics, through $G(0)$, where the
theory is inadequate; especially in field theories of condensed
matter, we {\em know} that the continuum description breaks down below
a lattice-spacing or an equivalent short length scale. But the effect
of the short distance physics on large length scale (low energy)
observables, such as the polarizabilities $C^{(\ell)}$, is to
renormalize (or ``dress'') their values. The values we obtained for
these polarizabilities, through a procedure that has nothing to do
with the short distance physics of the field, are therefore
renormalized values. On the other hand, if a field theory is to make
finite predictions on (low energy) physical observables, the
renormalized couplings must be accompanied by counter-terms, or
equivalently the coupling constants must be restored to their ``bare''
values. However, since the counter-terms exist for the sole purpose of
canceling these ultraviolet divergences in the values for physical
observables, it is legitimate not to explicitly write the
counter-terms, but discard such divergences on these grounds when
encountered, as we did in this article. We will nevertheless provide
an example below.

For the sake of simplicity, let us assume two particles possessing
only a dipole polarizability described by the Hamiltonian
\begin{equation}
  \Delta \HH = \half C_1 \ff^2_i (\bfr_1) + \half C_2 \ff^2_i (\bfr_2)
  \ .
\end{equation}
It is not hard to see that the divergences in the first and second
cumulants, $\cml{ \beta \Delta \HH}$ and $-(1/2) \cml{ (\beta \Delta
  \HH)^2}$, are of the trivial sort, \ie\ those that do not
involve the distance between particles. However, in the third cumulant
we encounter a divergent {\em pair} free energy
\begin{equation}
  \frac{1}{3!} \binom{3}{2} \frac{C_1^2 C_2} { (2\sigma)^3} 8
  G^{11}_{ij} G^{12}_{jk} G^{21}_{ki} \ , \label{eq:divgPair}
\end{equation}
as well as a similar term with the labels 1 and 2 interchanged. Using
the fact that $\del_i \del_j G(0) = (1/2) \delta_{ij} \nabla^2 G(0)$,
we may rewrite (\ref{eq:divgPair}) as
\begin{equation}
  \half \left[ \frac{C_1^2 } {2 \sigma^2} \nabla^2 G(0) \right]
  \frac{C_2}{\sigma} G_{jk}^{12} G_{kj}^{21} \ , \label{eq:divgPair2}
\end{equation}
where it is apparent that this energy scales as $\delta(0)/r^4$. The
way the factors were arranged in (\ref{eq:divgPair2}) makes it easy to
identify what counter-term should be added to the Hamiltonian to
negate this unphysical divergence: One might check that the {\em
  physical} pair energy between these two particles at the second
cumulant has exactly the form of (\ref{eq:divgPair2}), except that the
factor in square brackets is $C_1/\sigma$. This means that if one adds
to $\Delta\HH$ the counter-term
\begin{equation}
  \HH_{\rm counter} = \half \frac{ C_1^2}{2 \sigma}\nabla^2G(0)
  \ff^2_i (\bfr_1) \ ,
\end{equation}
it produces a divergent term in the second cumulant that exactly
cancels the aforementioned divergence, (\ref{eq:divgPair}) or
(\ref{eq:divgPair2}), in the third cumulant. In fact, this
counter-term removes all divergences caused by a single occurrence of
$ G^{11}_{ij}$ in the entire cumulant expansion , but it is beyond the
scope of this article to prove this. Also note that adding this
counter-term to the Hamiltonian is equivalent to redefining the dipole
polarizability $C_1$ as
\begin{equation}
  C_1 \rightarrow C_1 \left( 1 + \frac{C_1}{2\sigma} \nabla^2 G(0)
  \right) \ .
\end{equation}

This was a simple example that exhibits an induced dipole charge that
produces a divergence due to interacting with itself once. Considering
other multipole orders will introduce new counter-terms involving
different divergences, such as $\nabla^4 G(0)$. Removing divergences
due to one charge interacting with itself more than once is achieved
by counter-terms of higher orders in the divergences, such as
$(\nabla^2 G(0))^2$. Regarding these counter-terms as restoring the
polarizabilities to their (divergent) bare values $C^{(\ell)}_{\rm
  bare}$ from the renormalized $C^{(\ell)}$, one may write
\begin{equation}
  C^{(\ell)}_{\rm bare} = C^{(\ell)} + \sum_{n=1}^{\infty}
  b^{(\ell)}_n \left[ \frac{C^{(\ell)}} {\sigma} \nabla^{2\ell} G(0)
    \right]^n \ ,
\end{equation}
where the coefficients $b^{(\ell)}_n$ are pure numbers appropriately
chosen to cancel the offending diagrams. In this appendix, for
instance, we have explicitly found out that $b^{(1)}_1 = 1/2$.

\section{Induced monopoles} \label{sec:monopole}

In our treatment we avoided induced monopole terms in the effective
theory, \ie\ terms of the form
\begin{equation}
  \half \sum_a C^{(0)}_a \ff^2 (\bfr_a) \ , \label{eq:monterms}
\end{equation}
since the field of these charges---each proportional to $\log \lvert
\bfr - \bfr_a \rvert$---violates square integrability of its
gradient in $\mathbb{R}^2$. Such terms would describe particles with
frozen vertical fluctuations. We will discuss a possible workaround
for this issue.

Consider the regularized free surface Hamiltonian
\begin{equation}
  \HH [\ff] = \half \sigma \int \dd^2 r \left( \ff_i^2 +
  \lambda^{-2} \ff^2 \right)
\end{equation}
which approaches the tension Hamiltonian (\ref{eq:H}) in the limit of
large $\lambda$. For an interface between two fluids subject to
gravity (horizontal in its unperturbed state), $\lambda$ is the
capillary length and is given by $\lambda = \sqrt{\sigma / g \rho}$,
where $g$ and $\rho$ are the gravitational acceleration and mass
density difference between the fluids, respectively. The addition of
the ``mass'' term damps correlations over distances larger than
$\lambda$, hence regularizing the infrared divergence of monopole
fields. The Green function for this choice of surface energy is
\begin{equation}
  G(\bfr, \bfr') = \frac{1}{2\pi} K_0 \left( \left \vert \bfr -\bfr'
  \right \vert / \lambda \right) \ ,
\end{equation}
instead of $(-1/2\pi) \log \left \vert \bfr - \bfr' \right \vert$,
where $K_0 (x)$ is a modified Bessel function of the second kind.
After this regularization, we may safely consider particles that are
completely pinned.

The effective theory of such particles involves the same induced
charges as before, \RF{eq:DeltaH}, with the addition of the monopole
($\ell=0$) terms (\ref{eq:monterms}). The reader could object that,
with the new choice of kernel, there may be new terms such as
$\ff_{ii}^2$ or $\ff_{ijkk} \ff_{ij}$, \etc., but these terms can be
eliminated using the equation of motion, $ (-\nabla^2 + \lambda^{-2})
\ff =0$ (for a proof, see \eg\ Ref.~\cite{TASI}). The matching of the
polarizability coefficients is done similarly by comparing the
response of the induced charges to backgrounds of the form $\ff^{\rm
  bg} = \alpha r^{\ell} \cos \ell \varphi$ in the full and effective
theories. One finds
\begin{equation}
  \delta \ff^{\rm full} (\bfr) = - \frac{ R^\ell} {K_\ell (R / \lambda )}
  K_\ell (r /\lambda ) \cos \ell \varphi
\end{equation}
and
\begin{equation}
  \delta \ff^{\rm eff} (\bfr) = - \frac {C^{\ell}} {\sigma} \frac{\ell!}
     {2\pi \lambda^\ell} K_\ell (r /\lambda) \cos \ell \varphi \ ,
\end{equation}
respectively, yielding
\begin{equation}
  C^{(\ell)} = \frac{ 2\pi R^\ell \lambda^\ell \sigma} {\ell! K_\ell
    (R /\lambda)} \ .
\end{equation}
Expanding the Bessel functions, one observes that for $\ell > 0$ these
polarizabilities converge to those in \RF{eq:C} as $\lambda
\rightarrow \infty$, whereas the massless limit of the monopole
polarizability is
\begin{equation}
  C^{(0)} \rightarrow - \frac{2 \pi \sigma} {\log \left( \dfrac {
      \ee^{\gamma} R} {2 \lambda} \right)} \quad \text{as } \lambda
  \rightarrow \infty\ ,
\end{equation}
where $\gamma$ is the Euler-Mascheroni constant.

We observe that the monopole polarizability vanishes like $1/ \log (R
/ \lambda)$ in the massless limit of infinite capillary length that we
eventually want to take. This means diagrams involving monopoles could
vanish as well, unless every factor of $1/ \log (R/\lambda)$ due to a
monopole polarizability is canceled by a similar factor in the
numerator. Factors of this form indeed exist: they come from
monopole-monopole links in the propagator product, since in the
massless limit $G(r) = -(1/2\pi) \log (\ee^\gamma r /2\lambda)$. Due
to the closed topology of the diagrams, there are enough
monopole-monopole links to balance the vanishing monopole
polarizabilities {\em only} when there are no higher order multipoles
in the diagram; replacing one monopole polarizability in the diagram
costs two monopole-monopole links. In other words, the only monopole
interactions that do not vanish in the massless limit are those with
other induced monopoles and nothing else. This elucidates and
generalizes the findings of Lehle and Oettel that monopole-dipole and
monopole-quadrupole interactions indeed vanish~\cite{LehOetPRE}.

We can now write the pair interaction between two pinned particles, on
a capillary surface for which the capillary length tends to
infinity. The interaction will consist of \RF{eq:fullasym} due to
induced multipoles of order $\ell > 0$ and, based on the discussion of
the previous paragraph, a part $\beta \FF^{\rm mon}$ that is due
solely to monopole polarizabilities. Observe that in the latter,
propagators do not carry any derivatives and therefore {\em all} the
cumulants are of the same order in the inter-particle
separation. Hence, all cumulants must be summed for the monopole
interactions. To evaluate this, one can refer to \RF{eq:pairdiags},
keeping in mind that there is no sum over multipole orders $\ell_i$
now; only the monopoles are taken. Assuming identical particle radii
$R$ to declutter expressions, one finds
\begin{align}
  \beta \FF^{\rm mon} =& -\half \sum_{s=1}^{\infty} \frac{1}{s} \left(
  \frac{ C^{(0)} G (r)} {\sigma} \right)^{2s} \nonumber \\ =& -\half
  \sum_{s=1} ^{\infty} \frac{1}{s} \left( \frac{ K_0 (r /\lambda)}
      {K_0 (R / \lambda)} \right)^{2s} \ .
\end{align}
Owing to the damping of the regularized theory, this series converges
for all $r\, (\ll \lambda \rightarrow \infty)$ to
\begin{align}
  \beta \FF^{\rm mon} =& \half \log \left[ 1- \frac{ K_0^2 (r /
      \lambda)} {K_0^2 (R/ \lambda)} \right] \nonumber \\ =& \half
  \log \left[ K_0 (R/ \lambda) -K_0 (r/ \lambda) \right] \nonumber
  \\ + &\half \log \left[ K_0 (R/ \lambda) +K_0 (r/ \lambda) \right] +
  \text{ const.}
\end{align}
When the massless limit $\lambda \rightarrow \infty$ is taken, the
first term on the right hand side of the last equality gives
\begin{equation}
  \beta \FF^{\rm mon} = \half \log \log \frac{r} {R}
  \ , \label{eq:monopoleint}
\end{equation}
in agreement with Ref.~\cite{LehOetPRE}. The second term is
proportional to $ \log \log (\lambda^2/Rr)$, which is associated with
a force $\sim 1/r \log (Rr/\lambda^2) \rightarrow 0$ as $\lambda
\rightarrow \infty$. We note that extension to particles of unequal
radii results in the change $R \rightarrow \sqrt{R_1 R_2}$ in
\RF{eq:monopoleint}.
\end{appendix}

\input{filmPaper.bbl}
%%\bibliography{refs}

\end{document}